\let\includefigures=\iffalse
%
\let\useblackboard=\iftrue
%
%
\newfam\black
\input harvmac.tex

\includefigures
\message{If you do not have epsf.tex (to include figures),}
\message{change the option at the top of the tex file.}
\input epsf
\def\figin{\epsfcheck\figin}\def\figins{\epsfcheck\figins}
\def\epsfcheck{\ifx\epsfbox\UnDeFiNeD
\message{(NO epsf.tex, FIGURES WILL BE IGNORED)}
\gdef\figin##1{\vskip2in}\gdef\figins##1{\hskip.5in}
\else\message{(FIGURES WILL BE INCLUDED)}%
\gdef\figin##1{##1}\gdef\figins##1{##1}\fi}
\def\DefWarn#1{}
\def\figinsert{\goodbreak\midinsert}
\def\ifig#1#2#3{\DefWarn#1\xdef#1{fig.~\the\figno}
\writedef{#1\leftbracket fig.\noexpand~\the\figno}%
\figinsert\figin{\centerline{#3}}\medskip\centerline{\vbox{\baselineskip12pt
\advance\hsize by -1truein\noindent\footnotefont{\bf Fig.~\the\figno:} #2}}
\bigskip\endinsert\global\advance\figno by1}
\else
\def\ifig#1#2#3{\xdef#1{fig.~\the\figno}
\writedef{#1\leftbracket fig.\noexpand~\the\figno}%
\global\advance\figno by1}
\fi
\useblackboard
\message{If you do not have msbm (blackboard bold) fonts,}
\message{change the option at the top of the tex file.}
\font\blackboard=msbm10 scaled \magstep1
\font\blackboards=msbm7
\font\blackboardss=msbm5
\textfont\black=\blackboard
\scriptfont\black=\blackboards
\scriptscriptfont\black=\blackboardss

\else

\fi
%
\def\yboxit#1#2{\vbox{\hrule height #1 \hbox{\vrule width #1
\vbox{#2}\vrule width #1 }\hrule height #1 }}
\def\fillbox#1{\hbox to #1{\vbox to #1{\vfil}\hfil}}
\def\ybox{{\lower 1.3pt \yboxit{0.4pt}{\fillbox{8pt}}\hskip-0.2pt}}
%
\def\d{\kappa}
\def\hSigma{{\hat\Sigma}}
\def\CY#1{{\rm CY$_{#1}$}}

\def\comments#1{}

\def\Nbar{{\bar N}}
\def\NRR{{N_{RR}}}
\def\NNS{{N_{NS}}}

\def\ib{{\bar i}}
\def\jb{{\bar j}}

\def\lb{{\bar l}}
\def\mb{{\bar m}}
\def\nb{{\bar n}}
\def\ab{{\bar a}}
\def\bb{{\bar b}}
\def\cb{{\bar c}}

\def\db{{\bar d}}
\def\Dbar{{\bar D}}
\def\zb{{\bar z}}
\def\Zb{{\bar Z}}

\def\taub{{\bar \tau}}

\def\p{\partial}

\def\half{{1\over 2}}

\def\Re{{\rm Re\hskip0.1em}}
\def\Im{{\rm Im\hskip0.1em}}

\def\vev#1{\langle{#1}\rangle}
\def\bigvev#1{\bigg\langle{#1}\bigg\rangle}

\def\cL{{\cal L}}
\def\cM{{\cal M}}

\def\II{\relax{I\kern-.10em I}}
\def\IIa{{\II}a}
\def\IIb{{\II}b}

\def\IZ{\relax\ifmmode\mathchoice
{\hbox{\cmss Z\kern-.4em Z}}{\hbox{\cmss Z\kern-.4em Z}}
{\lower.9pt\hbox{\cmsss Z\kern-.4em Z}}
{\lower1.2pt\hbox{\cmsss Z\kern-.4em Z}}\else{\cmss Z\kern-.4em
Z}\fi}
\def\IB{\relax{\rm I\kern-.18em B}}
\def\IC{{\relax\hbox{$\inbar\kern-.3em{\rm C}$}}}
\def\ID{\relax{\rm I\kern-.18em D}}
\def\IE{\relax{\rm I\kern-.18em E}}
\def\IF{\relax{\rm I\kern-.18em F}}
\def\IG{\relax\hbox{$\inbar\kern-.3em{\rm G}$}}
\def\IGa{\relax\hbox{${\rm I}\kern-.18em\Gamma$}}
\def\IH{\relax{\rm I\kern-.18em H}}
\def\II{\relax{\rm I\kern-.18em I}}
\def\IK{\relax{\rm I\kern-.18em K}}
\def\IN{\relax{\rm I\kern-.18em N}}
\def\IP{\relax{\rm I\kern-.18em P}}

%
\def\inbar{\,\vrule height1.5ex width.4pt depth0pt}

\def\p{\partial}

\def\pb{{\bar \p}}

\font\cmss=cmss10 \font\cmsss=cmss10 at 7pt
\def\IR{\relax{\rm I\kern-.18em R}}
\def\pbar{\bar{\p}}

\def\vol{{\rm vol}}

\def\CM{{\cal M}}
\def\CN{{\cal N}}
\def\CZ{{\cal Z}}
\def\BR{\IR}
\def\BZ{\IZ} 
\def\BP{\IP}
\def\BR{\IR}
\def\BC{\IC}

\def\lp10{l_P^{10}}
\def\lp11{l_P^{11}}
\def\R11{R_{11}}
\Title{\vbox{\baselineskip12pt\hbox{hep-th/0307049}
\hbox{RUNHETC-2003-21}}}
{\vbox{
\centerline{Counting Flux Vacua}}}
\smallskip
\centerline{Sujay K. Ashok and Michael R. Douglas\footnote{$^\&$}{
Louis Michel Professor}}
\medskip
\centerline{Department of Physics and Astronomy}
\centerline{Rutgers University }
\centerline{Piscataway, NJ 08855--0849}
\medskip
\centerline{$^\&$I.H.E.S., Le Bois-Marie, Bures-sur-Yvette, 91440 France}
\medskip
\centerline{\tt ashok@physics.rutgers.edu, mrd@physics.rutgers.edu}
\bigskip
\noindent
We develop a technique for computing expected numbers of vacua in
Gaussian ensembles of supergravity theories, and apply it to
derive an asymptotic formula
for the index counting all flux supersymmetric 
vacua with signs
in Calabi-Yau compactification of type \IIb\ string theory, which
becomes exact in the limit of a large number of fluxes.
This should give a reasonable estimate for actual numbers of vacua
in string theory, for CY's with small $b_3$.

\Date{July 2003}
\def\np{{\it Nucl. Phys.}}

\def\pr{{\it Phys. Rev.}}
\def\pl{{\it Phys. Lett.}}
\def\atmp{{\it Adv. Theor. Math. Phys.}}

\def\cmp{{\it Comm. Math. Phys.}}
\nref\achflux{B. Acharya, [arXiv:hep-th/0212294].}
\nref\bagwit{J. Bagger and E. Witten, Phys. Lett. 115B (1982) 202.}
\nref\modcosmology{T.~Banks, M.~Berkooz, S.~H.~Shenker, G.~W.~Moore 
and P.~J.~Steinhardt, ``Modular cosmology,'' 
\pr\ D {\bf 52}, 3548 (1995) [arXiv:hep-th/9503114].}
\nref\beckerG{K. Becker and M. Becker,
\np\ B477 (1996) 155-167; [arXiv:hep-th/9605053].}
\nref\zelditch{P. Bleher, B. Shiffman and S. Zelditch, 
\cmp\ {\bf 208} (2000), 771.}
\nref\sz{P.~Bleher, B.~Shiffman, S.~Zelditch, 
{\it Invent. Math.} {\bf 142} (2000), 351-395 [arxiv:math-ph/9904020].}
\nref\boupol{R.~Bousso and J.~Polchinski,
``Quantization of four-form fluxes and dynamical 
neutralization of the  cosmological constant,''
JHEP {\bf 0006}, 006 (2000)
[arXiv:hep-th/0004134].}
\nref\candelas{P.~Candelas, X.~C.~De La Ossa, P.~S.~Green and L.~Parkes,
``A Pair Of Calabi-Yau Manifolds As An Exactly Soluble Superconformal  
Theory,''
\np\ B {\bf 359}, 21 (1991).}
\nref\cardoso{G.~L.~Cardoso, G.~Curio, G.~Dall'Agata, D.~Lust, 
P.~Manousselis and G.~Zoupanos,
``Non-Kaehler string backgrounds and their five torsion classes,''
\np\ B {\bf 652}, 5 (2003)
[arXiv:hep-th/0211118].}
\nref\dasgupta{K.  Dasgupta, G. Rajesh and S. Sethi,
JHEP 9908 (1999) 023, [arXiv:hep-th/9908088].}
\nref\dd{F. Denef and M.~R.~Douglas, to appear.}
\nref\dmw{
D.-E. Diaconescu, G. Moore and E. Witten,
``A Derivation of K-Theory from M-Theory,''
[arXiv:hep-th/0005091].}
\nref\jhstalk{M. R. Douglas, Lecture at JHS60, October 2001, Caltech.
Available on the web at {\tt http://theory.caltech.edu}.}
\nref\stat{M.~R.~Douglas,
``The statistics of string / M theory vacua,''
JHEP {\bf 0305}, 046 (2003)
[arXiv:hep-th/0303194].}
\nref\mrdtalks{M. R. Douglas, talks at Strings 2003 and at the
2003 Durham workshop on String Phenomenology.}
\nref\dsz{M. R. Douglas, B. Shiffman and S. Zelditch, work in progress.}
\nref\frey{A. R. Frey and J. Polchinski,
``N=3 Warped Compactifications,''
Phys.Rev. D65 (2002) 126009; hep-th/0201029.}
\nref\gkp{S.~B.~Giddings, S.~Kachru and J.~Polchinski,
 ``Hierarchies from fluxes in string compactifications,'' 
\pr\ D {\bf 66}, 106006 (2002)
[arXiv:hep-th/0105097] }
\nref\gv{R.~Gopakumar and C.~Vafa,
``On the gauge theory/geometry correspondence,''
\atmp\  {\bf 3}, 1415 (1999) [arXiv:hep-th/9811131].}
\nref\gp{M.~Grana and J.~Polchinski,
``Gauge / gravity duals with holomorphic dilaton,''
\pr\ D {\bf 65}, 126005 (2002)
[arXiv:hep-th/0106014].}
\nref\cstring{B.~R.~Greene, A.~D.~Shapere, C.~Vafa and S.~T.~Yau,
``Stringy Cosmic Strings And Noncompact Calabi-Yau Manifolds,''
Nucl.\ Phys.\ B {\bf 337}, 1 (1990).}
\nref\gvw{S.~Gukov, C.~Vafa and E.~Witten, ``CFT's from Calabi-Yau 
four-folds,''
Nucl.\ Phys.\ B {\bf 584}, 69 (2000) [Erratum-ibid.\ B {\bf 608}, 477
(2001)] [arXiv:hep-th/9906070].}
\nref\gukov{S.~Gukov,
``Solitons, superpotentials and calibrations,''
\np\ B {\bf 574}, 169 (2000) [arXiv:hep-th/9911011].}
\nref\gurrieri{S.~Gurrieri, J.~Louis, A.~Micu and D.~Waldram,
``Mirror symmetry in generalized Calabi-Yau compactifications,''
Nucl.\ Phys.\ B {\bf 654}, 61 (2003)
[arXiv:hep-th/0211102].}
\nref\HePG{
A.~M.~He and P.~Candelas,
``On The Number Of Complete Intersection Calabi-Yau Manifolds,''
Commun.\ Math.\ Phys.\  {\bf 135}, 193 (1990).}
\nref\horne{J.~H.~Horne and G.~W.~Moore, ``Chaotic coupling constants,''
\np\ B {\bf 432}, 109 (1994) [arXiv:hep-th/9403058].}
\nref\kklt{S.~Kachru, R.~Kallosh, A.~Linde and S.~P.~Trivedi,
``De Sitter vacua in string theory,'' [arXiv:hep-th/0301240].}
\nref\kst{S.~Kachru, M.~B.~Schulz and S.~Trivedi,
``Moduli stabilization from fluxes in a simple IIB orientifold,''
[arXiv:hep-th/0201028].}
\nref\kachru{S.~Kachru, M.~B.~Schulz, P.~K.~Tripathy and S.~P.~Trivedi,
``New supersymmetric string compactifications,''
JHEP {\bf 0303}, 061 (2003)
[arXiv:hep-th/0211182].}
\nref\gmoore{G.~W.~Moore, 
``Arithmetic and attractors,'' [arXiv:hep-th/9807087].}
\nref\moorec{G.~W.~Moore, private communication.}
\nref\psflux{J. Polchinski and A. Strominger,
\pl\ B388 (1996) 736-742, [arXiv:hep-th/9510227].}
\nref\polnytimes{J. Polchinski, as quoted by D. Overbye,
{\it The New York Times}, Oct 29, 2002.}
\nref\strom{A.~Strominger, ``Superstrings with Torsion'', 
\np\.B274 (1986) 253.}
\nref\stromspec{A.~Strominger, ``Special Geometry,''
\cmp\ {\bf 133} (1990) 163.}
\nref\wb{{\it Supersymmetry And Supergravity}, J.~Wess and J.~Bagger}
\nref\siegel{{\it Symplectic Geometry}, C. L. Siegel, 1964, Academic Press,
New York.}
\nref\tritri{P.~K.~Tripathy and S.~P.~Trivedi,
``Compactification with flux on K3 and tori,''
JHEP {\bf 0303}, 028 (2003)
[arXiv:hep-th/0301139].}
\nref\zelreview{S.~Zelditch, Proceedings of ICM, Beijing 2002, {\bf 2}, 
733--742, [arxiv:math.CA/0208104].}
%
%

\newsec{Introduction}

Among the many variations on string and M theory compactification, one
of the simplest is to turn on $p$-form field strengths in the internal
(compactification) space.  First studied in \strom, these ``flux
vacua'' have received a lot of recent attention, because it is
relatively easy to compute the flux contribution to the effective
potential, in terms of an exact superpotential which displays a lot of
interesting physics: it is dual to nonperturbatively generated gauge
theory superpotentials, it can stabilize moduli, it can lead to
spontaneous supersymmetry breaking, and it may be central in
explaining the smallness of the cosmological constant.  Out of the
large body of work on this subject, some important and representative
examples include 
\refs{\achflux,\beckerG,\boupol,\cardoso,\dasgupta,\frey,%
\gkp,\gp,\gurrieri,\kklt,
\kst,\kachru,\psflux,\tritri}.

In this work, we study the number and distribution of flux vacua in
Calabi-Yau compactification of type \II\ string theory.  We give an
explicit formula for an ``index'' counting all supersymmetric flux
vacua with signs, as an integral over configuration space, using
techniques which generalize to a large class of similar problems.  One
can start from any similar ensemble of flux superpotentials, and one
can get similar (though more complicated) explicit formulas for the
total number of supersymmetric vacua, for the index and number of
stable nonsupersymmetric vacua, and even for the resulting
distribution of supersymmetry breaking scales and cosmological
constants.  We defer detailed exploration of these generalizations to
\refs{\dsz,\dd}\ and future work, but use their general form to argue that the
index we compute is a reasonable estimate for the total number of
supersymmetric vacua, and even for the total number of metastable
non-supersymmetric vacua.

We review the basic definitions in sections 2 and 3.
The basic data of a flux vacuum in a theory compactified on $M$ is a
choice of flux, mathematically an element of $H^p(M,\BZ)$.\foot{This
is an oversimplification, as is explained in \dmw, but will suffice
for our purposes.}  It can be parameterized by the integrals of the
field strength on a basis of $p$-cycles, call these $\vec N$.
In $\CN=1$ supersymmetric compactification, the flux superpotential \gvw\ is
linear in the flux $\vec N$,
\eqn\firstW{
W(z) = \vec N \cdot \vec \Pi(z) .
}
Here $\vec\Pi(z)$ are contributions from individual fluxes, which
can be found as central charges of BPS domain walls \gukov.
In some examples, such as Calabi-Yau compactification
of the type \II\ string, the $\vec\Pi(z)$ are explicitly computable,
using techniques developed in the study of mirror symmetry \candelas.
In \IIb\ compactification, 
one takes $p=3$ and the $\Pi(z)$ are periods
of the holomorphic three-form.  In the mirror \IIa\ picture, the same
results can be thought of as incorporating world-sheet instanton corrections.

One can argue that the K\"ahler potential is independent of the
flux, in which case it is determined by $\CN=2$ supersymmetry and
special geometry.  The result is a completely explicit formula for the
scalar potential, which includes many (though not all) world-sheet and
space-time non-perturbative effects.  Almost always, the result is a
complicated and fairly generic function of the moduli $z$, which has
isolated critical points, in physical terms stabilizing all moduli
which appear explicitly in \firstW.

The resulting set of vacua is further reduced by identifications
following from duality.  An example in which this is simple to see is
compactification on $T^6$ or a $T^6/\BZ_2$ orientifold \refs{\kst,\frey}, 
in which
case the relevant duality is the geometric duality $SL(6,\BZ)$.\foot{
While there is a larger T-duality group, it does not identify
Calabi-Yau compactifications, but produces new, non-K\"ahler
compactifications
\refs{\kachru,\gurrieri,\cardoso}}.  
In general Calabi-Yau compactification, duality makes
identifications $(z,\vec N) \sim (z',\vec N')$, and we should factor
this out.  This can be done by restricting $z$ to a fundamental region
in the moduli space, after which any two choices $\vec N\ne \vec N'$
will lead to distinct flux vacua.

To the extent that one can choose $\vec N$ arbitrarily, the choice of
flux appears to lead to a large multiplicity of vacua, perhaps
infinite.  The first to try to quantify this were Bousso and Polchinski
\boupol, who suggested that a large 
number of flux vacua, say $N_{vac} \sim 10^{120}$, might provide a
solution of the cosmological constant problem, by leading to a
``discretuum'' of closely spaced possible values of $\Lambda$
including the observed small value $\Lambda \sim 10^{-120} M_{pl}^4$.
They went on to argue that the number of flux vacua should go as
$N_{vac} \sim L^K$, where $K$ is the number of cycles supporting flux,
and $L$ is an ``average number of fluxes,'' which in their argument
depends on an assumed ``bare negative cosmological constant.''  Since
a typical Calabi-Yau threefold has $K\sim 100$, such an estimate would
make a large number of vacua very plausible.

While numbers like $10^{120}$ vacua may seem outlandish, from a
broader point of view they just reinforce the point, which emerged
long ago from study of the heterotic string on Calabi-Yau
(see for example \HePG), that string
and M theory compactification involves many choices.  At present we
can only guess at the number of possibilities, and serious attempts to
characterize and come to grips with this aspect of the theory are only
beginning.

As emphasized in \refs{\jhstalk,\stat}, it is very important to bound
the number of string vacua which resemble the Standard Model and our
universe, because if this number is infinite, it is likely that
string/M theory will have little or no predictive power.  Going
further, this observation can be made quantitative, as was proposed in
\stat, by developing estimates for the number of vacua meeting one or
several of the tests for agreement with real world physics, such as
matching the scales and hierarchies, the gauge group, properties of
the matter spectrum, supersymmetry breaking and so forth.  As
explained there, such estimates can tell us how predictive we should
expect string/M theory to be, and provide a ``stringy'' idea of
naturalness.  Making useful estimates requires having some control
over each aspect of the problem, in particular we need to know why the
number of flux vacua is finite and get a controlled estimate 
of this number, with upper and lower bounds.

As was appropriate for an exploratory work, Bousso and Polchinski's
arguments were heuristic, and it was not obvious how to turn them into
any sort of controlled estimate for numbers of vacua; in particular
they did not take back reaction or duality into account.  This is the
problem which we address in the present work, and in some cases solve,
providing an estimate for the number of supersymmetric vacua which
becomes exact as the ``number of fluxes'' (to be defined shortly)
becomes large, using techniques which can provide precise bounds and
generalize to a wide variety of similar problems.  Although the
details differ from \boupol, the results confirm the suggestion that
numbers of flux vacua grow as $L^K$, and determine the overall
coefficient.

We now discuss the specific problem we treat in a bit more detail.
If no conditions are put on $\vec N$, the number of vacua is
infinite, because the problem of finding solutions of $DW=0$ or $V'=0$
is independent of the scale of $W$.  If one places a positive definite
condition on $\vec N$, such as $|N|<N_{max}$, then the number of
allowed values of $\vec N$ is finite, and finiteness of the total
number of vacua will follow if for each given $\vec N$ the number of
flux vacua is finite.  Since CY moduli spaces are compact, this is
plausible {\it a priori}, and we will verify it below.

However, it is not obvious why there should be such a bound on the
flux.  In Bousso and Polchinski's treatment, one assumed each flux
made an $O(1)$ positive contribution to the cosmological constant, so
fixing the cosmological constant led to such a bound.  However, this
assumption is not obviously true after taking back reaction into account.

In the case of type \II\ compactification on orientifolds, 
as discussed by Giddings, Kachru and Polchinski \gkp, tadpole
cancellation leads to a condition \refs{\gvw,\gkp}\ which sets the
scale of $\vec N$ as
\eqn\tadpoleone{
\eta(N,N) = L ,
}
In itself, this is not a bound on the flux,
since $\eta$ is an {\it indefinite} quadratic form, but one can also
argue (as we review later) that $\eta(N,N)>0$ for supersymmetric vacua.
However, this is still not enough to force the number of vacua to be
finite; indeed,  infinite series of supersymmetric flux vacua
in compactification on $T^6/\BZ_2$ and $K3\times T^2$  orientifolds
were found by Trivedi and Tripathy  \tritri\foot{
According to our definitions; see section 3.}
Fortunately, the infinite series they find does not spoil
predictivity, because it runs off to large complex structure, which
amounts to a partial decompactification limit.  Thus, all but a finite
number of these vacua are not really four dimensional.  However, this
example shows that the problem of finiteness is a bit subtle.

We will show that finiteness is true if we restrict attention to a
compact region of moduli space in which a non-degeneracy
condition is satisfied.  We believe this condition will fail only in
decompactification limits, in which case this result implies
that four dimensional supersymmetric flux vacua are finite in number.

Compared to the original problem of counting flux vacua,
our main simplification will be to ignore
the quantization of flux, instead computing the volume 
\eqn\defvol{
{\rm vol}\ R_{susy} = \int_{R_{susy}} d^K{\vec N} ,
}
of the region $R_{susy}$ in``charge space'' in which supersymmetric vacua
lie.  We will be more precise about this in section 3, but these words
give the right idea.  The intuition for why this should estimate the
actual number of flux vacua is very simple.  Flux vacua are points in
$R$, whose coordinates $\vec N$ are integers.  If one considers a
``reasonably simple'' region $R$, it is plausible that the number of
lattice points it contains, will be roughly its volume, and that this
will become exact in the limit of
large $L$.  However, there are subtleties which we 
will discuss.  Our tentative conclusion will be that this
is reasonable if $L>>K$, but may run into difficulties if $1<<L<<K$.

Another simplification, which is less essential, is to compute this
estimate for the ``supergravity index,'' which counts vacua with
signs.  Our techniques apply to both the index and to actual numbers of
vacua, but the simplest results are obtained for the index.  Of
course, the index is a lower bound on the total number of
supersymmetric vacua.  One can get moderately simple upper bounds as
well.

In fact our results will be somewhat more precise: we will work at
a point in configuration space, and compute an ``index density'' 
$d\mu_I(z)$ and ``vacuum density,''
$d\mu_{susy}(z)$
which measures the contribution to \defvol\ of a given point $z$ in
configuration space.  The total volume and thus the total estimated
number of vacua can then be obtained as an integral over
a fundamental region $\CF$ of moduli space,
\eqn\defdens{
{\rm vol}\ R_{susy} = \int_\CF\ d\mu_{susy}(z)
}

Having outlined the problem, we introduce our techniques for solving
it in section 4.  These were inspired by mathematical work on
counting zeroes and critical points of random sections of line bundles
\refs{\zelditch,\zelreview}.  While this work is fairly recent, and
the application to supergravity is new, the general ideas are fairly
well known in physics, especially in the study of disordered systems.
This will allow us to make our discussion self-contained and
non-mathematical, for better or worse.  We refer to
\dsz\ for a discussion of this problem and related problems in a more
mathematical language and for rigorous results.

Our basic technique is to reformulate the problem of computing the 
volume, as an expectation value in a Gaussian ensemble of superpotentials.
All expectation values in such an ensemble are determined by
a ``two-point function'' for the random superpotential,
$$
\bigvev{W(z_1) W^*(\zb_2)} =
 {1\over\CN} \int DW e^{-Q(W,W^*)} W(z_1) W^*(\zb_2) .
$$
where $Q$ is a quadratic form (the covariance), and $\CN$
is the overall normalization.
For many ensembles of interest, including
the flux ensemble, this turns out to be
$$
\bigvev{W(z_1) W^*(\zb_2)} = e^{-K(z_1,\zb_2)} ,
$$
where $K(z_1,\zb_2)$ is the standard K\"ahler potential on
moduli space, regarded as an independent function of the
holomorphic and antiholomorphic moduli.  This reduces all questions
about the distribution of flux vacua to geometric questions about
the moduli space.

The main result we derive
here is an explicit formula for the index density in such an ensemble,\foot{
Our conventions are given in section 2.  In these conventions,
the $1/n!$ factor which appeared in this and subsequent formulas in v1
is not present.}
$$
d\mu_I(z) = {1\over\pi^n }\det(-R-\omega) ,
$$
where $\omega$ and $R$ are the K\"ahler form and curvature for the
K\"ahler metric on configuration space at the point $z$.
We also discuss similar formulas for the total number of vacua of
various types, at least to the extent of arguing that they produce
similar results.  Although we will not do it here, one can also use
these techniques to study non-supersymmetric vacua, and to compute
expectation values which depend on the superpotential at several
points in configuration space, as will be discussed in \dsz.

In section 5, we apply these results to the specific case of \IIb\ flux
superpotentials, and make some simple physical observations.
The final result, for the index of all
supersymmetric vacua satisfying \tadpoleone, is
\eqn\sresult{
I_{vac}(L\le L_{max}) = \sum_{L\le L_{max}} I_{vac}(L)
 = {(2\pi L_{max})^{K}\over\pi^{n+1} K!}\int_{\CF\times\CH} \det(-R-\omega) ,
}
where $\CF$ is a fundamental region in Calabi-Yau moduli space,
and $\CH$ is a fundamental region of $SL(2,\BZ)$ in dilaton-axion
moduli space.
Techniques exist to work out this integrand
explicitly, so this is a fairly concrete
result, which could be evaluated numerically on a computer.

The primary observation is that in generic regions of moduli space,
the integral \sresult\ is closely related to the volume of the moduli
space.  Neglecting the curvature dependence, we might say that ``each
flux sector gives rise to one vacuum per $(\pi M_{pl}^2)^n$ scale
volume in configuration space.''

These volumes are in general believed to be finite \horne.  
Granting this claim, we answer our basic question, and show that the
number of physical flux vacua is finite.  This argument could fail near points
of diverging curvature; as an example, we discuss the conifold point
and find that the number of vacua near it is finite as well.

For $K>>L$, the formula \sresult\ predicts essentially no vacua.
We believe this is incorrect and merely shows that the discreteness
of the fluxes cannot be ignored in this case.  One can get a suggestive
estimate by taking into account the possibility that some fluxes vanish
by hand.

Although explicit volumes of moduli spaces have not been computed
for any physical \CY3 examples, they are known for simplified examples
such as tori with diagonal period matrix, or abelian varieties.  The
mathematical problem of finding flux vacua is perfectly well defined
in these cases and thus we can give precise results, which it would be
interesting to check by other means.

As a final comment, it would be quite interesting if a direct
topological field theory computation could be made of the index
counting supersymmetric vacua, perhaps by inventing some sort of
topologically twisted supergravity theory.

\newsec{Background, and ensembles of superpotentials}

The set of $\CN=1$ supergravity Lagrangians obtained by considering
the Gukov-Vafa-Witten superpotentials \gvw\ %
associated to all choices of flux in type \II\
compactification on Calabi-Yau, is an ensemble of effective field
theories, as defined in \stat.  

For many purposes -- testing the formalism, providing solvable
examples, studying universality claims and discussing to what extent
these approximate effective Lagrangians represent the exact situation
in string/M theory, it is useful to introduce and discuss more general
ensembles.  Thus we begin by reviewing the supergravity formula for
the effective potential, and defining the basic ensembles we will
consider.

\subsec{$\CN=1$ supergravity Lagrangian}

The data of an $\CN=1$ supergravity theory which concerns us is the
configuration space $\CC$, a complex K\"ahler manifold with K\"ahler
potential $K$, and a superpotential $W$.  We denote the complex dimension
of $\CC$ as $n$.

In general, we follow the conventions of \wb, with one exception -- we 
take the superpotential to be a section of a line bundle $\CL$ with
$$
c_1(\CL) = {\d\over\pi} \omega 
$$
where $\d$ is a real constant.
In supergravity, $\d=-1/M_{pl}^2$,
and in the body of the paper, we will set $M_{pl}=1$, so $\d=-1$.
However, all definitions entering into the effective potential
can be generalized to arbitrary $\d$, and this allows us to
discuss some similar and instructive problems.

Other than this generalization,
the rest of this subsection is review of standard definitions.
To define the line bundle $\CL$ over $\CC$, one works in patches.
In each patch, 
the K\"ahler potential is a function $K_{(a)}(z,\zb)$
satisfying $K(\zb,z)=K(z,\zb)^*$.  It determines
a Hermitian metric on configuration space,
\eqn\confmetric{
g_{i\jb} \equiv {\p^2 K \over \p z^i\p \zb^\jb} ,
}
which enters the kinetic term for the matter fields.  We also write
$$
\omega = {i\over 2} g_{i\jb} dz^i d\zb^\jb
$$ 
for the K\"ahler form, and 
$$
\vol_\omega = {1\over n!}\omega^n
$$
for the associated volume form.

The Riemann and Ricci curvatures for a K\"ahler manifold are
$$\eqalign{
R_{i\jb k}^l &= -\p_\jb (g^{l\mb}\p_i g_{k\mb}) \cr
R_{i\jb} &= R_{i\jb k}^k = -\partial_i\partial_{\jb}\log(\det g) .
}$$

The K\"ahler potentials in two overlapping patches $a$ and $b$ will be
related as
$$
K_{(a)} = K_{(b)} + F_{(ab)} + F_{(ab)}^*
$$
where $F_{(ab)}(z)$ is a holomorphic function (with
mass dimension 2) on the overlap.

This structure also defines an associated
holomorphic line bundle $\CL$ on $\cM$.
A section $\chi$ of $\CL$ is given by holomorphic functions $\chi_{(a)}$ in
each patch satisfying the condition
$$
\chi_{(a)} = e^{\d ~F_{(ab)} } \chi_{(b)} .
$$
This structure is preserved by the holomorphic
``K\"ahler-Weyl'' transformations
\eqn\translaw{\eqalign{
K \rightarrow K + f(z) + f^*(\zb) \cr
\chi \rightarrow e^{\d ~f} \chi .
}}
In general, $f(z)$ can be a different holomorphic function $f_{(a)}$
in each patch, in which case 
$F_{(ab)} \rightarrow F_{(ab)} - \d (f_{(a)} - f_{(b)})$, etc.\foot{
If $H_2(\CC,\BZ)$ is non-trivial, then
for the line bundle $\CL$ to be well defined,
$\d$ must be quantized so that $[\d\omega]\in H^2(M,\BZ)$.
This will come up in some of our toy examples.
It was also proposed long ago that this would be required
in supergravity \bagwit.  However, there are 
loopholes in the argument for this, as we discuss below.}

Given sections $\chi$ and $\psi$ of $\cL$, 
one can define the hermitian inner product
\eqn\invprod{
(\psi,\chi) \equiv e^{-\d~ K} \psi^*\chi
}
and the covariant derivative 
\eqn\covdergen{\eqalign{
D_i \chi &= \p_i \chi - \d~(\p_i K) \chi ;\qquad
\Dbar_\ib \chi = \pb_\ib \chi \cr
\Dbar_\ib \chi^* &= \pb_\ib \chi^* - \d~(\pb_\ib K) \chi^* ;\qquad
D_i \chi^* = \p_i \chi^* .
}}
The derivative $D_i \chi$ 
transforms as a section of $\cL \otimes \Omega\CM$,\foot{
$\Omega\CM$ is the bundle of $(1,0)$-forms.}
but in general is not holomorphic.  We also define
\eqn\invformprod{
(D\psi,D\chi) \equiv e^{-\d~K} g^{i\jb} (D_\jb\psi^*)(D_i\chi)
}
and so on.

The curvature of this connection (covariant derivative) is
$$\eqalign{
{i\over 2}[\Dbar_\jb,D_i] &= {i\d\over 2}\pb_\jb \p_i K \cr
 &= \d\omega .
}$$
In particular, the first Chern class of $\CL$ is 
$c_1(\CL)={\d\over\pi}[\omega]$.
Since $\omega$ is necessarily a positive hermitian form, the sign here
has important consequences.  In the supergravity case, $\CL$ is a
negative line bundle.

We take the superpotential $W$ to be a section of $\cL$.
It enters into the potential as
\eqn\potential{
V = (DW,DW) - {3\over M_{pl}^2} (W,W) 
 = e^{-\d~K}\left(g^{i\jb} (D_i W) (D_\jb W^*)
  - {3\over M_{pl}^2} |W|^2 \right).
}

We will consider ensembles in which $\CC$ and $K$ are fixed, while $W$
is taken from a distribution.  To get started, we might consider the
simplest possible choices for $\CC$ and $K$.  These are complex
homogeneous spaces, such as $\BP^n$, $\BC^n$ or $\CH^n$, the
$n$-dimensional complex hyperbolic space.

\subsec{Gaussian ensembles of superpotentials}

The primary ensemble we will treat is to take the superpotential as a
complex linear combination of sections of $\CL$, with a Gaussian weight.
We will eventually treat the physical flux problem as a limit of this.

Let $\Pi_\alpha$ with $1\le\alpha\le K$
be the basis of sections, then
\eqn\Wdist{
d\mu[W] =
\int \prod_{\alpha=1}^K d^2N^\alpha\ e^{-Q_{\alpha\beta} N^\alpha \bar N^\beta}
 \delta(W - \sum_\alpha N^\alpha \Pi_\alpha) .
}

Here $Q_{\alpha\beta}$ is a quadratic form (the covariance),
and $\bar N^\beta$ denotes the complex conjugate of $N^\beta$.
One could instead take real $N^\alpha$; the complex case
is slightly simpler and will turn out to be a better analog of the
\IIb\ string flux superpotential.  

We denote an expected value in this ensemble as
$$
\bigvev{ X } = {1\over\CN}\int d\mu[W]\ X ,
$$
where $\CN$ is an appropriate normalization factor.  For a unit
normalized ensemble,
$$
\CN = \int d\mu[W] = {\pi^K\over\det Q} .
$$
Through most of the discussion, we will use this convention, but eventually
will switch to discuss the ensemble of flux vacua, which is
normalized to the total number of flux vacua.

If $X$ is polynomial in $W$ and $W^*$, such expected values
can be easily computed using Wick's theorem and the two-point function
\eqn\defG{
G(z_1,\bar z_2) 
 = (Q^{-1})^{\beta\alpha} \Pi_\alpha(z_1) \Pi^*_\beta(\bar z_2) .
}
For example,
$$
\bigvev{W(z_1)\ W^*(\bar z_2)} = G(z_1,\bar z_2) .
$$

The primary expectation value of interest for us will be the
index density for supersymmetric vacua,
$$
d\mu_I(z) = \bigvev{\delta^{2n}(DW(z)) \det D^2W(z) } ,
$$
to be computed in section 4.

\subsec{Example of $\CC=\BP^n$.}

This is a good example for test purposes.  Also, as a compact space,
it is easier to work with mathematically, as we discuss in \dsz.

We start with homogeneous coordinates $Z^i$ with $0\le i\le n$,
and go to inhomogeneous coordinates:
set $Z^0=1$ and use $z^i=Z^i/Z^0$ with $1\le i\le n$.  The K\"ahler
potential is then
\eqn\cpnK{
K = \log (1 + \sum_i |z_i|^2) \equiv \log (1+|z|)^2 \equiv \log (Z,\Zb)
}
while the metric is
$$
g_{i\jb} = {(1+|z|^2) \delta_{ij} - z_i\zb_j  \over (1+|z|^2)^2} .
$$
The Ricci and Riemann curvatures are
\eqn\curvatures{\eqalign{
R_{i\jb} &= 
-\partial_i\partial_{\jb}\log\left({1\over(1+\sum_{k}|z_k|^2)^{n+1}}\right)
 = (n+1)g_{i\jb} \cr
R_{i\jb k\lb} &= g_{i\jb}g_{k\lb} + g_{i\lb}g_{k\jb} .
}}

As mentioned earlier, we will let the superpotential $W(z)$ take
values in the line bundle $\CL=\CO_{\BP^n}(\d)$ of degree $\d$, such
that $c_1(\CL)=\d \omega/\pi$ (here $\d$ must be integer).  Sections of
$\CO_{\BP^n}(\d)$ are degree $\d$ homogeneous polynomials.  One could
write a basis $\Pi_\alpha$ for these polynomials and compute \defG\
for a general covariance $Q$.

Since $\CC$ is compact, a natural choice for
$Q$ is the inner product of sections in the hermitian metric on $\CL$,
\eqn\symcov{
Q_{\alpha\beta} N^\alpha \bar N^\beta = 
\int_\CC (vol_\omega)\ e^{-\d K} |N\cdot\Pi|^2 .
}
In this case, the covariance $Q$ will 
respect all the symmetry of $K$, and so will the two-point function
$G$.

Using \symcov\ to define the covariance for $\BP^n$,
the resulting two-point function \defG\ must be a $U(n+1)$-invariant
polynomial of bidegree $(\d,\d)$ in $Z_1$ and $\Zb_2$.  This determines
it to be
$$
G(Z_1,\Zb_2) = (Z_1,\Zb_2)^{\d}
$$
so
$$
G(z_1,\zb_2) = (1 + z_1\cdot \zb_2)^{\d} .
$$

Note that this can also be written as
\eqn\GdefK{
G(z_1,\zb_2) = e^{\d~ K(z_1,\zb_2)}
}
with $K$ as in \cpnK, reinterpreted by taking the holomorphic $z$
dependence a function of $z_1$ and the antiholomorphic $\zb$ dependence
a function of $\zb_2$.  This substitution can be made more precise
by using the formula
$$
K(z_1,\zb_2) = \sum_{m,n\ge 0} \left({z_1^m \zb_2^n\over m!n!} \right)
{\p^{m+n} K(z,\zb)\over \p^m z\p^n\zb}|_{z=\zb=0}
$$
which tells us that (given appropriate conditions) the function
$K(z,\zb)$ on $\CC$ determines the bi-holomorphic functions
$K(z_1,\zb_2)$ and $\exp \d K(z_1,\zb_2)$ on $\CC\times\CC$.

Since $\d>0$, this is not a supergravity ensemble.
One could instead take $\d<0$ and use a basis of sections
of $\CO(\d)$ with poles, to get toy supergravity examples.

\subsec{Example of $\CH^n$.}

Complex hyperbolic space appears as a supergravity configuration space
for compactification on homogeneous spaces, and can be regarded as
the ``trivial'' case of the special geometry we discuss shortly,
in which the Yukawa couplings are zero.

We use the coordinates $z^i$, $1\le i\le n$, and let $\CH^n$ be the
region $\sum_i |z_i|^2 < 1$, with K\"ahler potential
$$
K = - \log \left( 1 - \sum_i |z_i|^2 \right) .
$$
This space is noncompact and has $U(n,1)$ symmetry.  Its curvature
tensors are given by \curvatures\ with an overall change of sign.

There is a natural $U(n,1)$-invariant two point function,
$$
G(z_1,\zb_2) = (1 - z_1\cdot\zb_2)^{-\d} = e^{\d K(z_1,\zb_2)} .
$$
which again corresponds to using a polynomial basis of sections.

\subsec{Calabi-Yau moduli spaces}

We consider a Calabi-Yau $M$, which for generality we take $k$ complex
dimensional.  The configuration space $\CC$ is then its moduli space
of complex structures $\CM_c(M)$, of complex dimension $n$.  Its 
K\"ahler metric can be found using ``special geometry''
\stromspec, while the
flux superpotential is a linear combination of periods of the
holomorphic $k$-form $\Omega$.

We briefly review the most important parts of this for our purposes.
We start by choosing a fixed basis $\Sigma_\alpha$
for the middle homology $H_k(M,\BZ)$, and a Poincar\'e dual basis
$\hSigma_\alpha$ for the middle cohomology $H^k(M,\BZ)$,
in which the intersection form
\eqn\intform{
\eta_{\alpha\beta} = \int_M \hSigma_\alpha \wedge \hSigma_\beta
}
is canonical: for odd $k$, a symplectic form
$$
\int \left(\matrix{\hSigma_{2a-1}\cr \hSigma_{2a}}\right)
 \wedge
\left(\matrix{\hSigma_{2b-1}& \hSigma_{2b}}\right) =
\left(\matrix{0& \delta_{a,b}\cr -\delta_{a,b}& 0}\right) ,
$$
and for even $k$ an indefinite symmetric form.
Call a normalized basis $\Sigma_\alpha$, with 
$1\le \alpha\le K\equiv b_k$ (for $k=3$, $b_3=2n+2$).

A choice of complex structure defines a Hodge decomposition
$$
H^k(M,\BC) = \oplus_{p+q=k} H^{(p,q)}(M,\BC) ,
$$
a decomposition of the middle cohomology into $(p,q)$ forms.  

The intersection form \intform\ pairs $(p,q)$ and $(q,p)$ forms.
In the case of primary interest here, threefolds with $H^1(M,\BR)=0$, 
on a given subspace this is definite with sign $(-1)^p$, i.e.
\eqn\intformsign{
0 < (-1)^p i^{-k} \int \alpha^{(p,q)} \wedge (\beta)^{(q,p)} .
}
More generally, this is true of the primitive forms (e.g. see \gvw).

One can show that already the choice of $H^{(k,0)}(M,\BC)$ subspace
determines the complex structure.  This choice determines a
holomorphic $(k,0)$-form $\Omega$ up to
overall normalization.  A choice $\Omega_z$ at each $z\in\CC$ defines
a section $s$ of a line bundle $\CL$ over $\CC$, and a preferred
metric in which the norm of the section is $1$,
\eqn\metCY{\eqalign{
1 = (s,s)|_z &= e^{K(z,\zb)}\ i^k\ \int_M \Omega_z\wedge\bar\Omega_z .
}}
An infinitesimal motion on $\CC$ will vary $\Omega$ by a sum of
$(k,0)$ and $(k-1,1)$ forms.  One can use \metCY\ to define a covariant
derivative,
\eqn\firstderiv{
D_i\Omega = \p_i \Omega + (\p_i K) \Omega ,
}
which acting on $\Omega$ produces a pure $(k-1,1)$ form.  
Using $\Omega$, one has an isomorphism from the $(k-1,1)$ forms 
to $H^1(M,TM)$, the deformations of complex structure, and this can be
used to show that
$$
-(D_i\Omega,\Dbar_\jb\bar\Omega) = \p_i\pb_\jb K = g_{i\jb} 
$$
is the Weil-Peterson metric on $\CC$, which is the metric deduced from
Kaluza-Klein compactification of the \IIb\ supergravity.

We define the normalized periods to be
$$
\Pi_\alpha = \int_{\Sigma_\alpha} \Omega.
$$
They are sections of $\CL$ as well.
In terms of these, we can write \metCY\ as
\eqn\CYK{
K(z,\zb) = -\log (i^k\eta^{\beta\alpha} \Pi_\alpha(z) \Pi^*_\beta(\zb)) .
}

As we discuss further in the next section,
the flux superpotential can be written in terms of the periods as
$$
W = N^\alpha \Pi_\alpha .
$$
It is a section of the line bundle $\CL$, so that $e^K|W|^2$ is independent
of the choice of $\Omega$.

The analog of \GdefK\ in this case is the two-point function
\eqn\specG{
G(z_1,\zb_2) = \bigvev{W(z_1) W^*(\zb_2)}
 = i^k\eta^{\beta\alpha} \Pi_\alpha(z_1) \Pi^*_\beta(\zb_2) =
 e^{-K(z_1,\zb_2)} .
}
We will also need the ``holomorphic two-point function''
(this terminology is appropriate if one takes $N$ real in \Wdist),
\eqn\specH{
H(z_1,z_2) 
 = \eta^{\alpha\beta}
 \Pi_\alpha(z_1)\Pi_\beta(z_2) .
}
For $k=3$, one can show that its leading term is cubic,
It is odd under $z_1\leftrightarrow z_2$, furthermore
$$
H(z_1,z_2) = \int \Omega(z_1)\wedge \Omega(z_2)
$$
and thus
$$
{\p\over\p z_2}H(z_1,z_2)|_{z_2=z_1} =
 \int \Omega(z_1)\wedge {\p\Omega(z_1)\over\p z_1} = 0
$$
since $\p\Omega \in H^{(3,0)} \oplus H^{(2,1)}$.

Thus, it has the expansion
\eqn\cubicH{
H(z_1,z_2) = {1\over 6}
 \CF_{ijk}(z_1) (z_2-z_1)^i (z_2-z_1)^j (z_2-z_1)^k + \ldots
}
where the $\CF_{ijk}$ are (by an old tradition going back to
the early work on heterotic string compactification) called the
``Yukawa couplings.''  One can show \stromspec\ that for
$k=3$ they actually determine the Riemann tensor:
\eqn\specRfor{
R_{a\bb c\db} = -g_{a\bb} g_{c\db}
 -g_{a\db} g_{c\bb} + e^{2K} \CF_{acm} \CF^*_{\bb\db\nb} g^{m\nb} .
}

Special geometry for Calabi-Yau three-folds has been much studied and
enjoys many additional properties, such as the existence of special
coordinates and the prepotential.  Furthermore, the techniques for
explicitly computing periods are highly developed, and numerous
examples are worked in the literature, starting with the quintic
\candelas.  We will quote a few of these results as we need them below.

\subsec{Example of $T^{2}$ moduli space.}

The simplest Calabi-Yau manifolds are complex tori.  We choose real
coordinates $x^i$ and $y^i$ with $1\le i\le k$, and periodically
identify $x^i \cong x^i+1$ and $y^i \cong y^i+1$.  The complex
structure will then be defined by the complex coordinates
$z^i = x^i + \sum_j Z^{ij} y^j$, where $Z^{ij}$ is a $k\times k$
complex matrix with positive definite imaginary part.

Thus, the moduli space of complex structures on $T^{2k}$ is the
space of complex $k\times k$ matrices $Z_{ij}$ with positive definite
imaginary part, subject to identifications under a $GL(2k,\BZ)$ duality
symmetry, which acts geometrically on the torus (see \kst\ for a
detailed discussion of this). The K\"ahler potential is
\eqn\torusK{
K = -\log \det \Im Z .
}
and has $SL(2k,\BR)$ symmetry.

A normalized basis of $H^k(M,\BZ)$ can be taken to be the $\left({2k
\atop k}\right)$ $k$-forms obtained by wedging $dx^i$ and $dy^i$.
Integrating the holomorphic $k$-form $\Omega=\wedge_{i=1}^k dz^i$ then
produces as periods, all the cofactors of the matrix Z.

For our purposes, all this can be summarized in the
two-point function associated to the covariance
$$
Q_{\alpha\beta} = i^{-k} \eta_{\alpha\beta} ,
$$
as
$$
G(Z_1,\Zb_2) = e^{-K(z_1,\zb_2)} = (2i)^{-n} \det (Z_1 - \Zb_2) .
$$

For $k=1$, this moduli space 
is equivalent to $\CH^1$, but in a different coordinate
system related as $z = (1+iZ)/(1-iZ)$.
One then has
$$g_{i\jb} = \delta_{i\jb}{1\over 4(\Im Z)^2}; \qquad
R_{i\jb} = -2 g_{i\jb} .
$$
The volume of the standard fundamental region,
$$
\CF=\{Z\in\BC:\Im Z>0\ {\rm and}\ |Z|\ge 1\ {\rm and}\ |\Re Z|<\half\} ,
$$
is
\eqn\ttwovol{
{\pi\over 12} = \int_\CF {d^2\tau\over 4(\Im\tau)^2} .
}

We note that this volume does not satisfy the quantization
condition discussed in subsection 2.1; the associated line bundle has
$c_1(\CL)=-1/12$.  This is not a mathematical contradiction as the
fundamental region $\CF$ is not a manifold; it has both cusp and
orbifold singularities.

Is this a physical contradiction?  This fundamental region is ubiquitous
as a supergravity configuration space; for example the \IIb\ dilaton-axion
takes values here in ten dimensions, and this descends to the four
dimensional compactifications of interest.  Apparently
\IIb\ supergravity violates the integer quantization,
and there is a loophole in the argument of \bagwit.

We believe the physics which allows this is
essentially that discussed in \cstring.  One might think that if $\CL$
were not quantized, observables constructed from
the fermionic fields (the gravitino and fermions in
chiral multiplets) would not be single-valued, which would lead to
contradictions.  However, to detect the
non-quantization of $\CL$, one must make bosonic field configurations which
explore an entire two-cycle in $\CC$.  An example would be a cosmic
string in four dimensions.  Such a background has curvature, and a
deficit angle at infinity proportional to the volume of the two-cycle.
In this background, the fermionic fields are 
single-valued, since they are sections of $\CL^{\pm 1/2}
\otimes S^\pm$, where $S^\pm$ are the spin bundles on space-time.

This argument seems to us to remove the need for the quantization
condition. Admittedly, we do not know the exact K\"ahler metric on
$\CC$, and one might consider other hypotheses; for example that
$\alpha'$ and $g_s$ corrections restore the quantization condition.
However, since the quantization condition clearly does not hold in
directly analogous examples with extended supersymmetry, there seems
no good reason to believe in it for $\CN=1$.

\subsec{Example of the Siegel upper half plane}

For $k>1$, it turns out that attempting to quotient by $SL(2k,\BZ)$
does not lead to a reasonable moduli space.  Rather, one must keep the
K\"ahler moduli as well, leading to the  Narain moduli space
$SO(k,k;\BZ)\backslash SO(k,k)/SO(k)\times SO(k)$.  This suggests that one must
keep the K\"ahler moduli to get a sensible result in this case.  Since
our main interest is in models to illustrate the Calabi-Yau case, we
do not pursue this here.

One way to get a simple toy model with only complex structure moduli
is to restrict attention to the complex tori with diagonal period
matrix.  The set of these is preserved by the subgroup
$SL(2,\BZ)^k \times S_k$, so the volume of the fundamental region is
\eqn\tdiagvol{
V_k = {1\over k!}\left({\pi\over 12}\right)^k .
}

Another way to restrict the problem to get a well-defined complex
structure moduli space, is to consider only the complex tori
with symmetric period matrix $Z_{ij}=Z_{ji}$.  These are known
as abelian varieties, because this is the subset of complex tori
which are projective varieties (can be embedded in some $\BP^n$).

This moduli space, the Siegel upper half plane,
has dimension $n=k(k+1)/2$.  Its K\"ahler potential
and metric are obtained by restriction from \torusK, while the duality
group is $Sp(2k,\BZ)$.  
Let $\CF_k$ be a fundamental region for this group.

The volume of $\CF_k$ was computed by Siegel \siegel;
it is\foot{
One needs to be careful about conventions.  Siegel's metric, (2) 
in \siegel, is $4$ times ours.  On the other hand, he absorbs a factor
of $2^{k(k-1)/2}$ in the volume element (top of p. 4).}
\eqn\siegelvol{\eqalign{
V_k &= 2^{1+k(k-1)/2-k(k+1)} \prod_{j=1}^k (j-1)! \zeta(2j) \pi^{-j} \cr
 &\sim \left({k\over2\pi}\right)^{k^2/2} .
}}
Its first few values are 
$V_1=\pi/12$, $V_2=\pi^3/8640$ and $V_3=\pi^6/65318400$.
While this grows very rapidly for large $k$,
because of the factors of $2\pi$ in the denominator, this 
asymptotic behavior sets in only for $k>30$.  

We can try to use this result as at least some indication of how these
volumes behave on Calabi-Yau three-folds.  For this purpose, it seems
reasonable to draw an analogy between $n=b_{2,1}(M)$, the number of
complex structure moduli of the CY $M$, and $k(k+1)/2$, the number of
complex structure moduli of the $k$-dimensional abelian varieties.
Granting this, and looking at the regime $n\le 480$, the volumes do
not become large.  Of course, this cannot be taken too seriously, as
general CY moduli spaces and duality groups might be quite different.

The Euler character is
$$
\chi_k = {(-1)^n\over\pi^n n!}\int_{\CF_k}
 \epsilon^{i_1\jb_1\cdots i_n\jb_n}
 \epsilon_{k_1\lb_1\cdots k_n\lb_n}
 R_{i_1\jb_1}^{k_1\lb_1} \cdots R_{i_n\jb_n}^{k_n\lb_n} 
$$
or equivalently
$$
\chi_k = {1\over\pi^n}\int_{\CF_k} \det (-R) ,
$$
where $R$ is the curvature two-form expressed as a hermitian
$n\times n$ matrix,
$$
R^l_k = R_{i\jb k}^l dz^i \wedge d\zb^\jb ,
$$
and $\det$ is the matrix determinant.
For example, 
$\chi_1 = 1/6$, since for $T^2$ we have $R=-2\omega$.
Again, this can be fractional, because the moduli space has orbifold
singularities.
We also quote \siegel: $\chi_2 = 1/720$ and $\chi_3 = 1/181440$.

\subsec{Summary}

We have argued in a variety of examples, which will include the
flux superpotentials of primary interest,
that there is a natural ensemble of
superpotentials in supergravity, characterized by the two-point function
\eqn\genG{
G(z_1,\zb_2) = \bigvev{W(z_1) W^*(\zb_2)} = e^{\d K(z_1,\zb_2)} .
}
The important point is that this is completely determined by the
K\"ahler potential, and thus all properties of this
ensemble are determined by the K\"ahler potential.
In section 4, we will make this explicit for
the ``index'' counting supersymmetric vacua with
signs, a similar index counting nonsupersymmetric vacua with
signs, and actual numbers of either type of vacua.

\newsec{Type \IIb\ compactification on \CY3 with flux}

We now discuss flux compactification of type \IIb\ string theory on a
\CY3 orientifold $M$ with fixed points and O3 planes, following \gkp.

For readers familiar with this discussion, let us first say that we
will simplify the problem, by totally ignoring the K\"ahler moduli of
$M$.  Our main reason for doing this is that the tree level effective
action is a bad guide to their physics, which is in fact controlled by
nonperturbative effects.  In fact, one can argue very generally that
these effects will break the ``no scale'' structure of the tree level
effective action and stabilize these moduli
\refs{\kklt,\stat}, leading to essentially the same physics we will
obtain by leaving them out.  On the other hand, there is no well
motivated ansatz for these nonperturbative corrections.  

It will become clear below that given the exact or even approximate
dependence of the effective action on the K\"ahler moduli, one could apply
our methods to count vacua in the full problem; for present purposes
little insight would be gained by making an ansatz for this here.
We will discuss the problem including the K\"ahler moduli elsewhere.

Thus, we take as configuration space $\CC = \CM_c(M)\times \CH$, where
$\CH$ is the space of values of the dilaton-axion $\tau = C^{(0)} +
i e^{-D}$.  As K\"ahler potential in the effective Lagrangian, we take
the zero flux K\"ahler potential, which is the sum of \CYK\ and
\torusK\ for $k=1$ with $Z_{11}=\tau$.
In principle, this could get additional flux dependent corrections,
but (as we sketch below) one can argue that these must vanish at large
volume.  Thus, in the spirit of our previous simplification, we ignore
this possibility.

Thus, we can base the discussion on the zero flux discussion and
corresponding $\CN=2$ supersymmetric Lagrangian.  The new feature is
the flux.  The underlying \IIb\ supergravity has two three-form field
strengths, the Ramond-Ramond field strength $F$ and the Neveu-Schwarz
field strength $H$.  These enter into the supersymmetry conditions and
all of the subsequent analysis, only in the  combination
\eqn\defG{
G = F + \tau H .
}
In a ground state, the equations of motion will
force $F$ and $H$ to be harmonic forms, which are thus determined in terms
of their periods on a basis of $3$-cycles,
\eqn\fluxN{
\NRR^\alpha = \eta^{\alpha\beta} \int_{\Sigma_\beta} F ; \qquad
\NNS^\alpha = \eta^{\alpha\beta} \int_{\Sigma_\beta} H .
}
These take quantized values which we denote $\NRR$ and $\NNS$.
They can be chosen arbitrarily
subject to one constraint:  the presence of a Chern-Simons term
$$
\int d^{10}x\ C^{(4)} \wedge F^{(3)} \wedge H^{(3)}
$$
in the \IIb\ Lagrangian modifies the
tadpole cancellation condition for the RR four-form potential, to
\eqn\tadpole{
\eta_{\alpha\beta} (\NRR)^\alpha (\NNS)^\beta = L ,
}
where $L$ is the total RR charge of O3 planes minus D3 branes. 
In supersymmetric vacua, one cannot have anti-D3 branes, so
$L$ is bounded above in supersymmetric vacua by the O3 charge.

The effective supergravity action in such a flux background is then
as above, with the superpotential \refs{\gvw,\gkp} 
\eqn\fluxW{
W = (\NRR + \tau \NNS) \cdot \Pi \equiv N \cdot \Pi
}
where we define 
$$
N \equiv \NRR+\tau \NNS; \qquad
\Nbar \equiv \NRR+\bar\tau \NNS .
$$
A very concise argument for this claim was given by
Gukov \gukov.  By wrapping a $(p,q)$ five-brane 
on a three-cycle, one obtains a BPS domain wall in four
dimensions, across which the flux $(F,H)$ jumps by $(p,q)$ units.
On the other hand, one can show that the domain wall tension is
precisely $\Delta W$, the variation of \fluxW\ (this argument is
simplified by a further reduction to two dimensions).  The BPS 
condition then implies that the superpotential is \fluxW, up to
a flux-independent constant.

This result can be confirmed by a direct ten dimensional analysis of
the supersymmetry conditions, as was done for Minkowski four
dimensions (zero cosmological constant) in \gp.  They found that
supersymmetry requires $G$ to be a primitive $(2,1)$ form.
The primitivity condition involves the K\"ahler form,
which we are ignoring.  To compare the rest, starting from \fluxW,
using \firstderiv\ one sees that
\eqn\twoonecond{
D_i W = 0 \leftrightarrow G^{(1,2)} = 0 ,
}
while one can also check that
\eqn\Dtaueqn{
D_\tau W = \NNS \cdot \Pi - {1\over\tau-\taub} W
 = -{\Nbar\cdot \Pi \over \tau-\taub} ,
}
so
\eqn\zerothreecond{
D_\tau W = 0 \leftrightarrow G^{(3,0)} = 0 .
}
Finally, the zero cosmological constant condition implies 
\eqn\threezerocond{
W=0 \leftrightarrow G^{(0,3)} = 0 .
}
Thus the supersymmetry conditions from the two arguments agree.
However the advantage of the supergravity argument is that it implies
the existence of the corresponding exact solution of \IIb\
supergravity, and thus these conditions must be exact at large volume.
This addresses the point raised at the start of the subsection, of
justifying the use of the zero flux K\"ahler potential.  Presumably, a
similar analysis for AdS$_4$ backgrounds with cosmological constant
(or no-scale nonsupersymmetric backgrounds)
would confirm this for $W\ne 0$ as well.

\subsec{Positivity and finiteness}

As mentioned in the introduction, one needs to put some condition
on the fluxes to have any hope that the total number of vacua will
be finite, simply because the condition $DW=0$ (as well as $V'=0$)
is independent of the overall scale of the flux.

Can \tadpole\ serve as this condition, or are additional conditions
required?  One suspects from the work of \tritri\foot{
In the revised version 3 of \tritri, it is argued that the
series found there violates the primitivity condition, and thus is
not an infinite series of vacua.  This involves
the K\"ahler moduli, which we are ignoring, so this example still
counts as an infinite series by the definitions here.
}
that we also need to
remove the large complex structure limit.  Does this suffice?

The main problem with \tadpole\ is that it controls an indefinite
quadratic form, and so has an infinite number of solutions.  On the
other hand, there is a more subtle positivity argument
given in \refs{\gvw,\gkp}, which shows that
\eqn\positivity{
0 < \eta_{\alpha\beta} \NRR^\alpha \NNS^\beta
}
for supersymmetric vacua.  In other words, if we had taken $L\le 0$
in \tadpole, we would find no supersymmetric vacua.

To see this, one uses the equality
\eqn\seepos{
\eta_{\alpha\beta} \NRR^\alpha \NNS^\beta = 
{i\over2\Im\tau} \int G \wedge G^* , 
}
the equivalences \twoonecond\ and \zerothreecond, which mean that
at a supersymmetric vacuum, $G\in H^{(2,1)} \oplus H^{(0,3)}$,
and \intformsign, which shows that \seepos\ is positive on this
subspace.

Evidently this does not imply that the number of vacua is finite; for
$L$ positive every solution of \tadpole\ obviously solves \positivity.

We now argue that it implies that in any infinite
sequence of vacua satisfying \tadpole,
all but finitely many must lie within a neighborhood
of a ``D-limit,'' meaning a point in the compactification of
$\CC$ at which the $n\times K$ matrix $D_i\Pi_\alpha$ is reduced in rank.
A large complex structure limit, in which some set of periods
$(\Pi_6,\Pi_4,\Pi_2,\Pi_0) \sim (\tau^3,\tau^2,\tau,1)$ as
$\Im\tau\rightarrow\infty$, is an example of a ``D-limit.''
Conifold and orbifold/Gepner points are
not; we do not know if there are others.

We want to use this to show that a sequence of distinct vacua (not
related by duality) must approach a large structure limit.  Now a general
sequence of vacua can stabilize moduli at a succession of points which
wander off in Teichmuller space (the cover of $\CC$ on which the periods
are single valued).  On the other hand, for any
sequence of vacua, we can use duality to find a corresponding sequence
in which the moduli sit entirely in a single fundamental region of the
moduli space $\CC$.  

Let us now consider a compact region $\CC$ within the fundamental region,
not containing ``D-limits.''  We now argue that this region can only
contain finitely many supersymmetric vacua satisfying \tadpole.
We do not want to assume vacua are isolated, so
we now consider a ``vacuum'' to be a connected 
component in $\CC$ of the solutions of $D_i W=0$ for a fixed flux $N$.

At a fixed point $Z$ in moduli space, the supersymmetry conditions
$DW = 0$ (or equivalently $G^{(1,2)}=G^{(0,3)}=0$)
define a linear subspace of ``charge space'' $H^3(M,\BC)$.
We just argued that $\eta$ is positive definite on this subspace;
therefore the set of vectors in this subspace satisfying
$\eta N \bar N \le 1$ is compact.  Taking the union of these sets
over all $Z\in \CC$, the resulting set can be seen to be compact,
and can thus enclose a finite number of lattice points, the quantized
fluxes which support supersymmetric vacua.  

The reason this argument does not prove finiteness is that the
supersymmetry conditions might change rank at some point $Z$, allowing
$\eta N \bar N$ to develop approximate null vectors near this point.
This is what happens in the example discussed by \tritri.  What we
have argued is that it can only happen in a ``D-limit.''  The known
example of a ``D-limit'' is in fact a decompactification limit (the
large volume limit of the mirror
\IIa\ theory).  If it is true that any D-limit is a decompactification
limit, then we have shown that the number of fluxes supporting
supersymmetric vacua (at fixed $L$) is finite after removing
decompactification limits.

To complete the argument, and show that the number of vacua is finite,
one would need to show that for a given flux, the solutions of $DW=0$
in $\CC$ form an algebraic variety (have finitely many components).
The reason this is true, is that the periods $\Pi(z)$ do not have
essential singularities, as is clear in explicit examples such as 
\candelas.  One could make more general arguments, but we shall not
attempt this here.

\subsec{Example of $T^2$}

All this may be too abstract for some readers' taste.  The example
of the $T^6/\BZ_2$ orientifold is discussed very concretely in
\refs{\kst,\tritri}.  Many of the features of the problem can
be seen by considering an even simpler toy example of ``fluxes on
$T^2$ with fixed dilaton''.

We consider the family of superpotentials on $T^2$ complex structure
moduli space,
$$
W = A Z + B
$$
with $A=a_1+ia_2$ and $B=b_1+ib_2$ 
each taking values in $\BZ+i\BZ$.  One then has
$$
DW=0 \leftrightarrow \Zb = -{B\over A} .
$$
A ``tadpole condition'' analogous to $\eta N\Nbar = L$ would be
$$
\Im A^* B = L .
$$

The simplest way to count these vacua is to use $SL(2,\BZ)$
invariance to set $a_2=0$, and allow solutions for any $Z$ satisfying
$\Im Z>0$.  This condition simply requires $L>0$.  We then have
$$
L = a_1 b_2
$$
which determines $a_1$.  The remaining $SL(2,\BZ)$ invariance can be
used to bring $b_1$ into the range $0\le b_1<a_1$.  Thus a vacuum is
given by a choice of integer $a_1$ dividing $L$, and a choice of $b_1$
which takes $|a_1|$ possible values.  We can furthermore take $a_1>0$,
taking into account $a_1<0$ by multiplying this result by $2$.

The result is that the number of vacua for given $L$ is
$$
N_{vac}(L) = 2 \sigma(L) = 2 \sum_{k|L} k ,
$$
where $\sigma(L)$ is a standard function discussed in textbooks
on number theory, with the asymptotics
$$
\sum_{L\le N} \sigma(L) =  {\pi^2\over 12}N^2 + \CO(N log N)
$$

Let us compare this with the volume in charge space which supports
supersymmetric vacua.  Again,
$DW=0$ is solved by $-\Zb = B/A$.  Changing variables to
$(\rho,A)$ with $B = \rho A$, one has
$$\eqalign{
\int d^2A\ d^2B\ \delta(L-\Im A^* B) 
&= \int d^2A\ d^2\rho\ |A|^2\ \delta(L-|A|^2 \Im \rho) \cr
&= \pi L \int {d^2\rho \over (\Im\rho)^2} .
}$$
Since the integrand is invariant under $\rho\rightarrow -\bar\rho$,
the constraints on the fundamental region for $Z$, translate 
to the same constraint on $\rho$.
Thus, the volume is
$\pi^2 L/3$. which agrees with the $L$
derivative of the previous computation.

This illustrates both how the tadpole cancellation condition leads to
a finite volume region in charge space, and that for large $L$ the
volume can be a good estimate of the number of vacua.  However this
direct approach is hard to carry out in general.

\subsec{Approximating the number of flux vacua by a volume}

We now discuss to what extent a sum over quantized fluxes, such as
$$
N_{vac} = \sum_{\NRR,\NNS\in\BZ} N_{vac}(\NRR,\NNS)
$$
can be approximated by an integral,
\eqn\substint{
\sum_{\NRR,\NNS\in\BZ} \rightarrow \int \prod_{i=1}^{2n+2} d\NRR_i\ d\NNS_i .
}
Since the equations $DW=0$ are independent of the overall scale of $N$,
one can scale $L$ out of the problem, and
$$
\sum_{{\NRR,\NNS\in\BZ\atop
\eta\NRR\NNS=L}} N_{vac}(\NRR,\NNS) =
\sum_{{\NRR',\NNS'\in\BZ/\sqrt{L}\atop\eta\NRR'\NNS'=1}}
  N_{vac}({\NRR'},{\NNS'}) .
$$
Thus, one expects the integral to give the leading
behavior for large $L$, meaning large compared to the other quantities
in the problem.  Two other quantities which clearly might become larger
are $K$, the number of fluxes, and $\Pi(z)$, the periods themselves,
in extreme limits of moduli space.  Thus these are the most obvious
potential sources of problems.

There are other subtleties as well.  For the sum to be well
approximated by an integral, the region in charge space containing
solutions must be of the same dimension as the charge space.  Thus,
this may not work well for overdetermined systems of equations, such
as $DW=W=0$ which describes supersymmetric Minkowski vacua.
Furthermore, the region should not contain ``tails'' whose width (in
any of the coordinates $N^\alpha$) runs off to zero for large $N$.
Here are some illustrative examples in two dimensions $(M,N)$.  In a
case like $M^2+N^2 < N_{max}^2$, for $N_{max}>>1$ the estimate is
quite good, and qualitatively not bad even for $N_{max}\sim 1$.  On
the other hand, in a case like $0 < MN < N_{max}$, the volume of the
region is infinite, while the number of lattice points it contains is
in fact finite.  Finally, for $N>0$ and $0 \le |M|N^2 < N_{max}$,
while the volume goes as $\int dN/N^2$ and is finite, the number of
lattice points is in fact infinite.

Thus, justifying this approximation requires detailed consideration of
the region in charge space containing supersymmetric vacua.  The
possibility that the volume diverges, while the number of vacua is
finite, is best excluded by showing that the volume is finite.  This
was checked directly by G. Moore \moorec\ in a related problem
(attractor points in the large complex structure limit of the
quintic \gmoore), and this result was some motivation for us to push through
the analysis of section 4, which provides formulas which can be
used to show finiteness.

More subtle problems might arise, if the boundaries of the region were
sufficiently complicated.  In light of our previous arguments, this
region can be described as follows: at a given $z\in\CC$, the
constraints $\vec{N}\cdot D\vec{\Pi}(z)=0$ determine a linear subspace of charge
space; the integral over $\CC$ takes a union of these subspaces, while
the constraint \tadpole\ can be reduced to a positive quadratic bound
on $N$.  This last condition is simple, while if we confine our
attention to supersymmetric vacua in the interior of $\CC$, one might
expect the resulting region to have relatively simple boundaries, with
bad behavior again associated to ``D-limits'' in which ratios of
periods are not bounded.  We are already removing ``D-limits'' from
$\CC$ as these vacua are unphysical, so this type of argument suggests
that there will be no problems of this type.  
This could be made more
precise, but we leave detailed considerations to future work.

It will actually turn out that, at least in the examples we study, the
total volume is finite, including the D-limits, and furthermore the
volume associated to D-limits is small.  This may be physically
significant, along the lines of \modcosmology.  It is also
mathematically convenient, because it means we do not have to specify
the cutoff, which would necessarily be somewhat arbitrary; the total
volume is also a good estimate for the number of physical vacua.

\subsec{Setup to compute volume of flux vacua}

We start by replacing the sum over fluxes by an integral, which can
also be thought of as a complex integral\foot{
Our convention is
$d^2N \equiv d(\Re N)d(\Im N) = (i/2)dNd\Nbar$.}
\eqn\inthatN{
(\Im\tau)^{-K} \int \prod_{\alpha=1}^K d^2N^\alpha .
}

To turn the problem into a computation in the Gaussian ensemble
\Wdist, we implement the condition \tadpole\ 
by using the Gaussian weight
\eqn\lpsum{\eqalign{
N_{vac}(\alpha)
 &= \sum_{vacua}
e^{-2\alpha(\Im\tau)\eta_{\alpha\beta} N_{RR}^\alpha N_{NS}^\beta} \cr
 &= \sum
e^{-i\alpha\eta_{\alpha\beta} N^\alpha \Nbar^\beta}
}}
and then doing a Laplace transform in $\alpha$,
\eqn\laplacetr{
N_{vac}(L\le L_{max}) \equiv \sum_{L\le L_{max}} N_{vac}(L) = 
{1\over 2\pi i}\int_C {d\alpha\over\alpha}\ e^{2\alpha(\Im\tau) L_{max}}\ 
N_{vac}(\alpha) .
}
Given \positivity, the sum \lpsum\ should converge for
$\Re\alpha > 0$, and given a reasonable $L$ dependence (it will
turn out to be power-like) can be continued to general $\alpha$.
The integral \laplacetr\ can then be done by closing the
contour with an arc at large $\Re\alpha < 0$.

Since the argument which led to \positivity\ was a bit subtle, we
will not assume it in making the computation, but instead see it come
out as follows.  We can cut off large flux with a positive
definite Gaussian, taking as covariance
\eqn\regcov{
Q_{\alpha\beta} = i\alpha \eta_{\alpha\beta} + \lambda \delta_{\alpha\beta} ,
}
and computing an $N_{vac}(\alpha,\lambda)$, in terms of
a two-point function
\eqn\specGlam{
G(z_1,\zb_2)|_{\alpha,\lambda} = 
{1\over\lambda^2-\alpha^2}\left( -\alpha e^{-K(z_1,\zb_2)}
 + \lambda \sum_\alpha \Pi_\alpha(z_1) \Pi^*_\alpha(\zb_2) \right) .
}

If we find we can continue $N_{vac}(\alpha,\lambda)$ from large
$\lambda>0$ to $\lambda\rightarrow 0$ along the real axis without
encountering divergences, this will justify the claim.
Assuming this works, the volume of flux vacua will be
given in terms of the two-point function at $\alpha=1$ and $\lambda=0$, 
as given in \specG.

\newsec{Expectation values in Gaussian ensembles}

We now discuss computation of expected numbers of vacua in a general
Gaussian ensemble of superpotentials.  Many further results of this type
can be found in \dsz.

\subsec{Expected supersymmetric index}

This counts vacua with the signs given by the fermion mass matrix, in other
words $\det D_iD_j W$.  We can express it as an integral of a density, the
expected index for supersymmetric vacua at the point $z$, which is
\eqn\susyindex{
d\mu_I(z) = \bigvev{\delta^{2n}(DW(z)) \det D^2W(z) } .
}
The determinant is present to produce a measure whose $z$ integral
counts each solution of $DW=0$ with weight $\pm 1$.
It is of the $2n\times 2n$ matrix 
\eqn\susydet{
D^2W = \left(\matrix{
\p_i D_j W& \p_i \Dbar_\jb W^*\cr
\pb_\ib D_j W& \pb_\ib \Dbar_\jb W^*}\right) .
}
At a critical point $DW=0$, it does not matter whether the outer 
derivative is covariantized.

The simplest computation of this density within our ensembles is in terms
of a constrained two-point function.
It can be computed by implementing the delta function constraints with
Lagrange multipliers in the integral, to obtain
$$\eqalign{
{1\over\CN}
\int &d\mu[W]\ \delta^n(DW(z_0))\delta^n(\bar DW^*(z_0)) \prod_i W(z_i)
 ~\prod_j W^*(\zb_j) \cr
&= \sum_\sigma \prod_i G_{z_0}(z_i,\zb_{\sigma \jb})
}$$
where 
\eqn\defZ{\eqalign{
\CN &= \int d\mu[W]\ \delta^n(D_i W)\delta^n(\bar D_{\jb}W^*) \cr
 &= {\pi^{K-n}\over \det Q }
\det_{a,\bar{b}} \vev{D_a W(z_0)~\bar D_{\bar{b}}W^*(z_0)}}
}
and
\eqn\defGzero{\eqalign{
G_{z_0}(z_1,\zb_2) &\equiv 
\bigvev{W(z_1)~W^*(\zb_2)}_{DW(z_0)=0} \cr
&= G(z_1,\zb_2) - ({\bar D_{\zb^a_0}}G(z_1,\zb_0)) ~
 (D_a {\bar D_{\bar b}}G(z_0,\zb_0))^{-1}~
 ({D_{z_0^b}}G(z_0,\zb_2)) ,
}}
which is easily checked to satisfy
$$
D_1 G_{z_0}(z_1,\zb_2)|_{z_1=z_0} =
\Dbar_2 G_{z_0}(z_1,\zb_2)|_{\zb_2=\zb_0} = 0 .
$$

In terms of this function,
\eqn\detmuI{
d\mu_I(z) = \det (D_{z_1^a} D_{z_1^b} \Dbar_{\zb_2^\cb} \Dbar_{\zb_2^\db}
 G_z(z_1,\zb_2))^n|_{z_1=z_2=z} .
}

For example, for $n=1$, we have
$$
d\mu_I(z) =
{1\over\pi}\left.{ D_1 D_1 \Dbar_2 \Dbar_2 G_0(z_1,\zb_2)
 - \Dbar_1 D_1 D_2 \Dbar_2 G_0(z_1,\zb_2) 
 \over D_1 \Dbar_2 G(z_1,\zb_2)}\right|_{z_1=z_2=z} .
$$

\subsec{Geometric computations}

We proceed to compute the coincidence limits of covariant derivatives of 
$G(z_1,\zb_2)$ and $G_z(z_1,\zb_2)$ which appeared above.

The first point to make, is that all quantities of the form
\eqn\defF{
F_{ab\ldots|mn\ldots}(z_0) \equiv
e^{-\d K(z,\zb)} (D_{1a} D_{1b} \ldots)
(\Dbar_{2m} \Dbar_{2n} \ldots) G(z_1,\zb_2)|_{z_1=z_2=z_0}
}
(resp. $G_z(z_1,\zb_2)$) are tensors constructed from
the K\"ahler form, curvature and its derivatives.  The K\"ahler
potential itself does not appear.

To see this, note that under the K\"ahler-Weyl transformation \translaw,
we have
$$
G(z_1,\zb_2) \rightarrow 
e^{\d f(z_1) + \d f^*(\zb_2)} G(z_1,\zb_2) 
$$
(resp. for $G_z(z_1,\zb_2)$).  In other words, as is obvious from
its definition \genG, $G$ transforms as a product of sections.
The covariant derivatives respect this law, and thus $F$ will be
a tensor.  Finally, all tensors which can be constructed from
derivatives of $K$ are of the stated form.

From this, it will follow that any ensemble observable defined at a
single point in $\CC$ (say the density of a given type of vacuum), or
as a single integral over $\CC$ (say the distribution of cosmological
constants), can be expressed in terms of the K\"ahler form, curvature
and its derivatives.

Let us proceed.  We start with \covdergen\ and
$$
G(z_1,\zb_2) = e^{\d K(z_1,\zb_2)} .
$$
Then
$$\eqalign{ & D_{1a} D_{2\bb}G(z_1,\zb_2) = D_{1a}\cdot \d\left({\p
 K(z_1,\zb_2)\over\p \zb_2^\bb}- {\p K(z_2,\zb_2)\over\p
 \zb_2^\bb}\right)e^{\d K(z_1,\zb_2)} \cr
 &= \d\left({\p^2 K(z_1,\zb_2)\over
 \p z_1^a\p \zb_2^{\bb}}+
\d\left({\p (K(z_1,\zb_2) - K(z_1,\zb_1))\over\p z_1^a}
{\p (K(z_1,\zb_2)- K(z_2,\zb_2))\over\p
 \zb_2^\bb}\right)\right)e^{\d K(z_1,\zb_2)} 
}$$
Thus,
\eqn\twoderiv{\eqalign{ 
D_{a} D_{\bb}G(z_0,\zb_0)
 &= \d\cdot {\p^2 K(z_0,\zb_0)\over \p z_0^a\p\zb_0^b}G(z_0,\zb_0) \cr 
&= \d\cdot g_{a\bb}\cdot G(z_0,\zb_0)
}}
and
$$
{1\over\CN}=\pi^{K-n}\d^{-n} e^{ -n \d K}(\det g)^{-1} .
$$
This determines \defGzero.
The calculation of $F_{ab\cb\db}$ is similar.  We get
\eqn\fone{\eqalign{
F_{ab|\cb\db} &= e^{-\d K(z_0,\zb_0)}
D_{1a} D_{1b} \bar D_{2c} \bar D_{2d}\
G_z(z_1,\zb_2)|_{z_1=z_0,\zb_2=\zb_0} \cr &= \d\cdot\left({\p^4
K(z_0,\zb_0)\over \p z_0^a \p z_0^b \p \zb_0^c \p \zb_0^d}-{\p^3
K(z_0,\zb_0)\over \p z_0^a \p z_0^b \p\zb_0^m}g^{m\nb}{\p^3
K(z_0,\zb_0)\over \p z_0^n \p\zb_0^c
\p\zb_0^d}+\d(g_{b\cb}g_{a\db}+g_{a\cb}g_{b\db})\right) \cr 
&= -\d R_{a\cb
b\db}+\d^2(g_{b\cb}g_{a\db}+g_{a\cb}g_{b\db})
}}
and
\eqn\ftwo{\eqalign{ 
F_{a\bb|c\db} &= e^{-\d K(z,\zb)}
D_{1a} \bar D_{1\bb} D_{2c} \bar D_{2\db}
G_z(z_1,\zb_2)|_{z_1=z_0,\zb_2=\zb_0} \cr
&= \d^2\ g_{a\bb}g_{c\db}
}}
Note that the combination \fone\ vanishes for $\BP^n$ with $\d=1$,
or $\CH^n$ with $\d=-1$.  For special geometry, using \specRfor, 
we have
$$
F_{ab|\cb\db} = e^{2K} \CF_{abm} \CF^*_{\cb\db\nb} g^{m\nb} ,
$$
Despite the negative curvature this is manifestly positive.

Finally, mixed correlators such as
$D_{1a} D_{1b} D_{2c} \Dbar_{2\db} G_z$ are zero, as there is
no geometric invariant with this index structure.

\subsec{Result for the index density}

The index density is now obtained by substituting
these results into \detmuI.  Let us do this for a unit normalized
ensemble.  The computation is most easily done by writing the
determinant as a Grassmann integral,\foot{
This is essentially the mass matrix for the fermions
in the chiral superfields in the original supergravity.}
$$
\det D^2W = \int \prod_{i} d^2\psi^i 
d^2\theta^i\ e^{\theta^a \psi^b \p_a D_b W + 
\bar\theta^\ab \psi^b \pb_\ab D_b W + {\rm c.c.}} .
$$
Evaluating this in the Gaussian ensemble produces
$$\eqalign{
\bigvev{\det D^2W} &=
{1\over\pi^n}
\int \prod_{i} d^2\psi^i 
d^2\theta^i\ e^{\theta^a \bar\theta^\cb \psi^b \bar\psi^\db
F_{ab\cb\db} + 
\theta^a \bar\theta^\db \psi^c \bar\psi^\bb 
F_{a\bb c\db} } \cr
&=
{1\over\pi^n}
\int \prod_{i} d^2\psi^i 
d^2\theta^i\ e^{-\d\theta^a \bar\theta^\cb \psi^b \bar\psi^\db
R_{a\cb b\db} 
+ \d^2 g_{a\cb} \theta^a \bar\theta^\cb g_{b\db} \psi^b \bar\psi^\db} .
}$$
since the term \ftwo\ cancels the ``cross term'' in \fone\ coupling
$g\theta\bar\psi$.

One can then introduce an orthonormal frame 
$e^i_\alpha \bar e^\jb_{\bar\alpha} \delta^{\alpha\bar\alpha}$ and
change variables $\theta^i\rightarrow e^i_\alpha \theta^\alpha$.
This produces a determinant which cancels the $\det g$ from
$\CZ$.  Thus one obtains
\eqn\dmuIresult{
d\mu_I(z) = {1\over \pi^n }\det( -R + \d \omega\cdot{\bf 1} ) 
}
where $R$ is the curvature two-form, acting as a
$k\times k$ matrix on an orthonormal basis for $\Omega\CM$,
and ${\bf 1}$ is the $k\times k$ unit matrix.
For example, in one dimension, it is
\eqn\dmuIone{
d\mu_I(z) = {-R + \d \omega \over \pi},
}
where $R$ is the curvature two-form.

A more conceptual way to see this, and to check the precise normalization,
is to observe that for $\CC$ compact
and $\d$ positive, topological considerations determine \dmuIresult\ up
to a possible total derivative.  
Its cohomology class must be
\eqn\dmuIclass{
[d\mu_I] = c_n(T^*\CC \otimes \CL) ,
}
the top Chern class of the bundle $T^*\CC \otimes \CL$ in
which $D_i W$ takes values.
The combination $-R+\d\omega$ appearing in \dmuIresult\ is precisely
the curvature of this bundle.
On the other hand, the direct computation we just described cannot produce
total derivative terms.  Thus \dmuIresult\ is the exact result.

While our result reproduces the natural density coming out of a much
simpler topological argument, conceptually it is rather different.
First of all, the topological argument gives the index for a single
superpotential, while we have computed the expected index for an
ensemble of superpotentials.  Thus we will be able to use our result
to compute a sum over flux sectors.  

Equally importantly, any given
flux superpotential is not single valued on a fundamental region
of the moduli space $\CC$.  If one follows a loop around a singularity,
it will undergo a monodromy to become a different superpotential,
appropriate for a value of the fluxes related by a duality.
To use the topological argument, one must go to a covering space on
which the superpotential is single valued.  However, such a covering
space will not be noncompact, and cannot be compactified (the upper
half plane is a good example).  Thus, one cannot interpret the
integral of \dmuIresult\ over a fundamental region as the index for a
single superpotential; indeed its value will not usually be an integer.

In our computation, \dmuIresult\ arises as an expected value for
an ensemble of superpotentials which is invariant under monodromy.
This is why it is well defined on a fundamental region, and why it
makes sense to integrate it over a fundamental region.

Finally, the topological argument cannot be generalized to other
quantities such as the actual numbers of vacua.  Let us proceed to do
this for our computations.

\subsec{Expected numbers of supersymmetric vacua}

We would now like to compare the index we just computed to the actual
numbers of supersymmetric vacua.  The obvious way to do this would be
to compute
\eqn\dmuvac{
d\mu_{vac}(z) = \bigvev{ \delta(DW(z)) |\det D^2W(z)|} .
}
Of course this integral can not be done by Wick's theorem, and
analytic results are more difficult.

The point which makes this feasible, is that we are still only doing
Gaussian integrals.  A convenient way to phrase the computation,
following the work of Shiffman and Zelditch \refs{\sz,\dsz}, is to define a
joint probability distribution for the random variables
\eqn\secondders{
\xi_{ij} = \p_i D_j W(z) ,\qquad
\xi'_{\ib j} = \pb_\ib D_j W(z) = g_{\ib j} W(z),
}
under the constraint $DW(z)=0$.  As we implicitly used in writing
\defGzero, this is a Gaussian distribution; for example
$$
F_{ab|\cb\db} =
{e^{-\d K(z,\zb)}}
\bigvev{ D_a D_b W(z) \Dbar_\cb\Dbar_\db W^*(\zb)}
$$
can be reproduced by the Gaussian distribution
$$
\int d^2\xi\ e^{-(F^{-1})^{ab\cb\db} \xi_{ab} \bar\xi_{\cb\db}} .
$$
Thus, expectation values of any function of $D^2W(z)$, including
the non-analytic function in \dmuvac, are functions of the geometric
data $F_{ab\cb\db}(z)$ and $F_{a\bb c\db}(z)$ we already computed,
which could be found explicitly by doing finite dimensional integrals.

Let us consider the case $n=1$.  We denote the random variables
\secondders\ as $\xi$ and $\xi'$.  We then need to compute the integral
$$
\CI = {1\over\pi^2}
\int d^2\xi d^2\xi'\ e^{-|\xi|^2-|\xi'|^2}
 |F_{11\bar 1\bar 1}|\xi|^2 - F_{1\bar 1 1\bar 1}|\xi'|^2| .
$$
This is non-analytic in $F$; for 
$F_{11\bar 1\bar 1}/F_{1\bar 1 1\bar 1} < 0$ it is
$$
\CI =  {1\over\kappa g_{1\bar 1}}
|F_{11\bar 1\bar 1} - F_{1\bar 1 1\bar 1}| ,
$$
while for
$F_{11\bar 1\bar 1}/F_{1\bar 1 1\bar 1} > 0$ it is
$$
\CI = {1\over\kappa g_{1\bar 1}}
{ F_{11\bar 1\bar 1}^2 + F_{1\bar 1 1\bar 1}^2 \over
|F_{11\bar 1\bar 1} + F_{1\bar 1 1\bar 1}| }.
$$

This is already a little complicated, and clearly the analogous
expressions for higher dimensions will be rather complicated.

However, a simple consequence of this which is surely true in
higher dimensions, is that if the curvature (say the holomorphic
sectional curvature) stays bounded, then the ratio of the total
number of vacua, to the index, will be bounded.  It is probably
most interesting to get an upper bound on the total number of 
vacua (since the index serves as a lower bound).  One way to do
this would be to use Hadamard's inequality, which applied to the
matrix at hand takes the form
$$
|\det D^2W| \le \prod_i (\sum_j |\xi_{ij}|^2 + |\xi'_{i\jb}|^2) .
$$
Using $\xi'_{i\jb}=g_{i\jb}W$ and $\vev{W^*(z) D^2W(z)}=0$, this
can be brought to a reasonably simple form.  In $k=1$ this gives
$\CI \le |F_{11\bar 1\bar 1}|+|F_{1\bar 1 1\bar 1}|$.

\subsec{Nonsupersymmetric vacua}

The most interesting quantity is the number of metastable
({\it i.e.} tachyon-free) vacua, given by
\eqn\Nnonsusy{
N_{ms}=\int_{\cal{C}} d\mu_{X}(z) =
 \int [d\mu(K,W)]\ \int_\CC\ [dz] \delta^{(2n)}(V') (\det V'')\ \theta(V'')
}
where $V$ is given by \potential, and $\theta(V'')$ is the constraint
that the matrix $V''$ is positive definite.  (As it stands,
this includes supersymmetric vacua as well.)
Of course there are simplifications
of this; for example by leaving out the $\theta(V'')$ but keeping the
signed determinant one would get a Morse-type index for $V$, counting
all vacua with signs.

Compared with the supergravity index, the main additional complication
in computing this index is that the condition $V'=0$ is quadratic in
the flux $N$.  This can be treated by Lagrange multipliers, in a way
similar to how we are treating the constraint \tadpole.  One can also
control other quantities with quadratic dependence on the flux this
way, such as the cosmological constant.  We postpone further
discussion to \dd.

The main point we make here, is that these results are also determined
in terms of local tensors constructed from the K\"ahler metric.  Since
$V'' \sim D^3W DW$, they can involve up to six derivatives of the
Green's function, which will bring in up to two derivatives of the
Riemann tensor.

\newsec{Application to counting \IIb\ flux vacua}

The formalism we set up can be applied directly to compute the index
of flux vacua, in the approximation where we take the fluxes
$\NRR+\tau \NNS$ to be general complex numbers.

In the full problem, one must also solve the equation $D_\tau W=0$.
Perhaps the most straightforward way to do this is to take $\Pi_i$ and
$\tau \Pi_i$ as the basis of periods.  We then need to redo the above
calculations taking the Gaussian integral over real fluxes.  This
should lead to the same topological density \dmuIresult, because we
can again argue by comparison to the case of compact $\CC$ (one still
needs to check that extra total derivative terms cannot appear).  If
so, the final result should be the same as before, taking 
$\CC=\CM_c(M)\times\CH$ as the
configuration space.  This computation will appear in \dd.

Here, we will reach the same result, by a shorter argument using special
features of the case at hand.

\subsec{Flux vacua at fixed $\tau$}

As discussed in section 3, to check whether the previous results are
appropriate for counting flux vacua, we need to redo the computations
with the two-point function
\specGlam, and study the $\lambda\rightarrow 0$ limit.

Now, assuming the periods $\Pi$ and their derivatives stay finite,
the only place where divergences can enter the final result is 
in the overall normalization of the Gaussian integral, \defZ.
Thus we need to compute \twoderiv\ using \specGlam.  
This is
\eqn\twoderivlam{
D_{a} D_{\bb}G(z,\zb)|_{\alpha,\lambda}
 =
{1\over\lambda^2-\alpha^2} \left(-\alpha \d~ g_{a\bb}~e^{\d K(z,\zb)}
 + \lambda \sum_\alpha M_{a,\alpha} M^*_{\bb,\alpha}  \right)
}
where
\eqn\defM{
M_{a,\alpha}(z) = D_a\Pi_\alpha(z) .
}
The prefactor $1/(\lambda^2-\alpha^2)$ cancels between numerator
and denominator in \detmuI, so it causes no problem.
Since the second term is a product $M M^\dagger$, it is a
non-negative and hermitian matrix.
Since $\d=-1$, $-\d g_{a\bb}$ is positive definite and hermitian as well.

Thus, for real $\alpha>0$,
\twoderivlam\ is a positive definite hermitian matrix.
Thus, in this case the integral is finite.  On the other hand, for
$\alpha<0$, \twoderivlam\ can have zeroes and the integral will
generally diverge.  This matches the expectations from section 3,
namely that since supersymmetric vacua satisfy \tadpole, the integral
should only be finite for $\alpha>0$.\foot{
Note that we did not yet enforce $D_\tau W=0$, so $G$ will have
a $(0,3)$-form component, and the quadratic form is still indefinite.
What this argument is actually showing is that the signature of $Q$
on the constrained subspace $DW=0$ does not change as we take 
$\lambda\rightarrow 0$.  Thus, carefully doing the resulting integral 
by analytic continuation should lead to an extra $i$ in the next two
formulas.  This is not relevant for the real problem
with $D_\tau W=0$ enforced.
}

Redoing the computation of subsection 4.4, with the correct normalizations,
produces
$$
\bigvev{\delta(DW)\det D^2W} =
{\pi^{K-n}(-1)^{K/2}\over(\alpha~ \Im\tau)^K} \det(-R-\omega) .
$$
Doing the Laplace transform then produces
\eqn\expectedI{
I_{vac\ (fixed\ \tau)}(L\le L_{max})
 = {(2\pi L_{max})^{K}(-1)^{K/2}\over\pi^{n} K!}\int_\CF \det(-R-\omega)
}
where the integral is taken over a fundamental region of the duality
group in $\CC$.

As the simplest check of this result, the $T^2$ result 
$\pi^2 L^2/6$ follows directly from \ttwovol.

\subsec{Treating the dilaton-axion $\tau$}

From \Dtaueqn, we can implement the $D_\tau W=0$ condition by taking
\eqn\susyindextau{
d\mu_I(z,\tau) = 4(\Im\tau)^2
\bigvev{\delta^{2n}(DW(z)) 
\delta^{(2)}(\Nbar\cdot \Pi)\ \det_{i,j,\tau,\taub} D^2W(z)}.
}
The prefactor arises from extracting $2\ \Im\tau$ from each
of the two constraints \Dtaueqn.  
The new constraint can be solved along the same lines as \defGzero.  
It leads to an additional factor $1/\pi G$ in $\CZ$.

We now need to compute an $(2n+2)\times(2n+2)$ determinant
of the form \susydet.  This contains terms as before, and new terms
$$\eqalign{
\vev{\pbar_\taub D_\tau W \p_\tau \Dbar_\tau W^*} = (g_{\tau\taub})^2 G; \cr
\vev{\p_\tau D_\tau W \pbar_\taub \Dbar_\tau W^*} = (g_{\tau\taub})^2 G; \cr
\vev{\p_\tau D_i W \pbar_\taub \Dbar_\jb W^*} = g_{\tau\taub} g_{i\jb} G .
}
$$
In fact, all of these terms are the same as would be obtained by
using the same formulae \fone,\ftwo\ for the $\tau$ 
derivatives, with the two-point function
$$
G(z_1,\tau_1;\zb_2,\bar\tau_2) = (\tau_1-\bar\tau_2)G(z_1,\zb_2) .
$$
Thus, one can follow the same reasoning which led to \dmuIresult,
to obtain the same formula, but now
taking as configuration space $\CC=\CM_c\times\CH$, and using the
direct product K\"ahler metric.

The only problem with this reasoning is that the $D_\tau W$ constraint
couples $\Nbar$ to $\Pi$.  This 
leads to corrections to the two-point function proportional to 
$H = \eta^{\alpha\beta}\Pi_\alpha\Pi_\beta$ as in \specH.
The resulting constrained two-point function is
$$
G^\tau_z(z_1,\zb_2) = 
 G_z(z_1,\zb_2) - e^{K(z,\zb)} H(z_1,z) H^*(\zb,\zb_2) .
$$
On the other hand, by virtue of \cubicH,
the new term in $G_\tau$ vanishes to fifth order in the
coincidence limit $z_1=z_2=z_0$, 
and hence does not contribute to \susyindex.

Thus, the expected index in this case, as suggested by
general arguments, is \expectedI\ modified as we just described,
to a form on the full configuration space.\foot{
This is similar to taking \expectedI\ multiplied by the volume of $T^2$
moduli space, as stated in v1, but includes additional mixed terms, as
we will see in the example below.  This simplification, that the 
holomorphic two-point function drops out, is also not shared by more
general computations, such as the actual number of vacua \dd.}

One could follow the same steps with the numbers of supersymmetric or
nonsupersymmetric vacua discussed in section 4.  Since the entire
covariance is proportional to $\alpha$, it scales out of all of the
integrals in the same way, and the Laplace transform works in the
same way.  So, we get precisely the same power-like $L$ dependence for
all of these quantities, multiplied by different geometric factors.

\subsec{Finiteness of the number of vacua}

The upshot of all of this, is a formula which we claim estimates the
number of physical ({\it i.e.} truly four dimensional) flux vacua
in \IIb\ orientifold compactification with fluxes, in terms of the
geometry of CY moduli space:
\eqn\finalI{
I_{vac}(L\le L_{max})
 = {(2\pi L_{max})^{K}(-1)^{K/2}\over\pi^{n+1} K!}\int_\CF \det(-R-\omega)
}
While there are many points in our arguments which could be refined,
it is already interesting to ask if the geometric quantity which
appears is finite.  Some time ago, Horne and Moore conjectured that
volumes of these moduli spaces are finite \horne, and pointed out
possible consequences of this for stringy cosmology.  Granting this,
the remaining issue is whether the curvature dependence can lead to
divergences.

An example in which the Riemann curvature diverges is the neighborhood
of a conifold point \candelas.  This point is at finite distance, it
is not a ``D-limit'' by our formal definition, and physically does not
correspond to decompactification.  Thus an infinite number of vacua
near this point would be a problem.

We quote results for the complex structure moduli space of the mirror
of the quintic CY
\candelas\ (of course conifold points on other CY's should have the same
behavior).  This is a one dimensional moduli space; we quote the
K\"ahler and curvature two-forms, in terms of a coordinate $z$ which
vanishes at the conifold point:
$$
\omega_{z\zb} = -a^2 \log |z|; \qquad
R_{z\zb} = {1\over 2a^2 |z|^2 (\log|z|)^2}
$$
(here $a$ is a known constant).

While $R$ is singular, it is integrable.  This is a little tricky:
changing variables as $z=\exp 2\pi i u$, one has 
$$
R_{u\bar u} = {1\over 2a^2 (\Im u)^2}
$$
which at first sight looks problematic.  On the other hand, 
the neighborhood of $z=0$ maps to $\Im u >> 0$ and $|\Re u|<\half$,
and the integral $\int R$ over this region is finite.

This looks very much like \ttwovol, and this is no coincidence.
One can explicitly count flux vacua for the superpotential
$$
W = A z + B (z \log z + {\rm const})
$$
along the same lines as we did for $T^2$.  In this case, 
one finds vacua at $u=-A/B + \CO(\exp -|A/B|)$, and imposing the
same conditions we did there leads to the same results.

Of course, with more moduli, there are many more complicated
degenerations, but on the strength of this example it is at least
reasonable to hope for finiteness more generally.

One might try to argue for finiteness in degeneration limits,
from the idea that the dual gauge theories at these singularities are
conventional gauge theories and must have finitely many vacua.  This
is probably true, but it is not clear to us how to make this precise.
One would still need to bring a condition like \tadpole\ into the
argument.  Also, the dual meaning of all of the flux parameters,
in particular the choice of NS flux in the examples of Gopakumar 
and Vafa \gv,  has never been fully explained.

\subsec{The case of $K>>L>>1$.}

Nothing seems to be known about total volumes or curvatures of Calabi-Yau
moduli spaces, so it is hard to know how important the geometric
factor is.  The examples of complex tori and abelian varieties
discussed in section 2 suggest that it is important, but
subdominant to the large factorial in the denominator, which rapidly
sends the volume to zero for $K>L$.  This was something of a surprise
to us, but in retrospect has a simple explanation.  Intuitively, one
can think of the computation (and particularly the Laplace transform
\lpsum) as summing over the various distributions of the total flux
$L$ among sets of cycles.  In some sense, the positivity bound
\positivity\ must hold not just for the total flux, but among pairs of
cycles as well.  Thus the integral over these distributions produces a
factor (the volume of a $K$-simplex) which falls off rapidly.

An example which illustrates this is to consider
$k$ copies of $T^2$, where we take as periods the $K=2k$ one-forms.
This can be done using the previous formalism
(note that $G\ne e^{-K}$ in this case; rather $G=\sum G_i$).
One can also do this directly:
distributing the flux $L$ among them, leads to
$$\eqalign{
N_{vac}(L\le L_{max}) &= \sum_{L_1+\cdots+L_k\le L} \prod N_{vac}(L_k) \cr
&\sim \int_0 \prod_{i=1}^k dL_i \ \theta(L-\sum_i L_i) 
\prod_{i=1}^k \left({\pi^2 L_i\over 3}\right) \cr
&\sim \left({\pi^2\over 3}\right)^k {L^{2k}\over (2k)!} \cr
&= {(4\pi)^k L^{2k} \over (2k)!} \vol(\CF)^k
}$$
in agreement with the above.

For this special case, since the flux on each $T^2$ must separately
satisfy $L_i\ge 1$, there are in fact no vacua for $K>2L$, and this
estimate is good.  However, in more general examples, which do not
factorize in the same way, there is no reason to expect any analogous
constraint.  We should say that we have not shown that there is not
such a constraint; rather that our estimate cannot be regarded as
evidence for it.

In general, while the volume does fall off for $K>L$, this is probably not
a good estimate for the actual number of vacua.  The most obvious
consistency condition we can test is 
$$
I_{vac}(L) < \sum_{L'\le L} I_{vac}(L') \sim {L\over K} I_{vac}(L) 
$$
if we consider the volume, but this is obviously false for the
actual number of vacua and $K\ge L$.

The problem is that for $K\sim L$, many cycles will have zero or
one flux, and the discreteness of the fluxes cannot be ignored.
Perhaps the simplest way to try to get a better estimate is by adding
counts of flux vacua with some subset of the fluxes set to zero by
hand.  This could be done by the same techniques, with the difference
that we leave some periods out of
\specG.  If vacua with these fluxes set to zero exist, they will show
up in the corresponding volume, while if they do not, either this
modified volume should be small, or else the consistency condition we
discussed at the start of the section should fail.  Some preliminary
study of this point suggests that this will not happen, so that each
possible choice of subset of $K-n$ fluxes to set
to zero will give a non-zero result.

If we assume this, and furthermore assume
that the geometric factor is at most exponential in $n$,
we find
\eqn\guess{\eqalign{
N_{vac} &\sim \sum_{n=1}^K \left({K\atop n}\right) 
{(2\pi c L)^n\over n!} \cr
&\sim e^{\sqrt{2\pi c K L}} \qquad {\rm for}\ K>>L.
}}
This is an intriguing and suggestive possibility, with hints
of a dual string description, but at the moment is no more than that.

\subsec{Numbers}

Our general conclusion is that \finalI\ is a good lower bound for the
number of flux supersymmetric vacua for $L >> K >> 1$, and that a
reasonable estimate for the total number is probably \finalI\
multiplied by a factor $c^K$ with some $c\sim 1$.  It would be
interesting to work out the leading corrections at large $L$, 
to develop a good method for the case $K>>L>>1$, 
and to check these results against other methods.

Let us do some low dimensional cases of diagonal tori and abelian varieties.
Because these moduli spaces are symmetric spaces, one has
$$
\det(-R-\omega\cdot{\bf 1}) = c\cdot\vol_\omega = C \cdot \det(-R)
$$ 
in terms of constant combinatoric factors $c$ or $C$, computable
at one point in moduli space.  

For the diagonal torus $T_D^{2k}$, $R=-2\omega$, and
$$\eqalign{
\det(-R-\omega\cdot{\bf 1}) &=
 \prod_{i=1}^k (2\omega_i - \sum_{j=1}^k \omega_j) \cr
&= c \cdot \prod_{i=1}^k \omega_i
}$$
with $c=-k!$ and $\omega_i$ the K\"ahler form of the $i$'th $T^2$.
Using \tdiagvol, this gives
$$
I_{vac} = (-1)^{K/2+1} {(2\pi L)^K \over (12)^k K!} .
$$
In this case, the dilaton-axion lives in another copy of $T^2$,
so the same result (with $k\rightarrow k+1$) can be used to count
vacua with arbitrary $\tau$.

For the abelian variety $T_A$,
it is simpler to express the integrand
using $C$ and $\det(-R)$, since the 
Euler character $\chi$ suffers less from normalization ambiguities.
Thus, we have
$$
I_{vac} = {(2\pi L)^K (-1)^{K/2}\over K!} \cdot C \cdot\chi_{\CF\times\CH} .
$$

We did the computation of $C$ for $T_A^6$ using computer algebra, obtaining
$$
\det(-R-\omega)|_{\CF} = {1\over 4}\det(-R)|_{\CF}
$$
and
$$
\det(-R-\omega)|_{\CF\times \CH} = {7\over 4}\det(-R)|_{\CF}\times\omega_\CH .
$$
which, combined with $\chi_3=1/181440$ from \siegel\ and the volume
$\pi/12$ of $\CH$, produce the expected index for vacua on $T^6/\BZ_2$
with symmetrized period matrix,
$$\eqalign{
I_{vac} &= {7\cdot (2\pi L)^{20}\over 4\cdot 181440\cdot 12\cdot 20!} \cr
&\sim 4\cdot 10^{21} \qquad {\rm for}\ L=32.
}$$

We suspect this is in fact a reasonable estimate for the number of
supersymmetric flux vacua on the $T^6/\BZ_2$ orientifold, but to really
make this precise; two more points should be discussed.

One issue, which does not arise for compactification on general Calabi-Yau,
is the primitivity condition: we should take this into account,
and remove the restriction to symmetric period matrices.  Since these
conditions are also linear, this is computable using the same techniques,
which would be a nice exercise.
We suspect this would lead to the same formula \finalI, integrated over
the full moduli space of metrics $SL(6,\BZ)\backslash GL(6,\BR)/SO(6,\BR)$.  

An issue of more general importance is that
we need to discuss stabilization of the K\"ahler moduli.
This is not very well understood at present, but (as discussed in
\stat) we would agree with the general arguments given by
Kachru, Kallosh, Linde and Trivedi \kklt\ that nonperturbative effects
should do this in many compactifications, and that this can be
understood in a controlled way at large volume and weak coupling.  
At present it remains
an important problem to find models in which this actually works,
but granting that it can, the question might arise as to what fraction of
these $4\cdot 10^{21}$ vacua sit at large volume and weak coupling, so
that their existence could be proven using present techniques.

The fraction at weak coupling follows from the
results here, which as we discussed give the distribution of vacua
in moduli space, including the dilaton-axion.  In particular, the number
of vacua with coupling $g_s^2 < \epsilon$, is obtained by simply restricting
the integral \ttwovol\ to this region in coupling space, giving
$$
I_{vac}|_{g_s^2 < \epsilon} \sim 3\int_{1\over\epsilon} {d\tau_2\over\tau_2^2}
\sim 3\epsilon I_{vac} .
$$
In other words, the distribution is essentially uniform at weak coupling,
so insisting on weak coupling does not dramatically reduce the number of 
vacua.

As discussed in \refs{\kklt,\stat}, the fraction which sit at large
volume is essentially the fraction with small $e^K|W|^2$ according to
the analysis here.  It turns out \mrdtalks\ that this distribution is
also uniform near zero, as will be shown in \dd, and this ``cut'' also
leaves large numbers of vacua.

It could still be that imposing further constraints, such as
acceptable supersymmetry breaking, or metastability, dramatically
lower the number of vacua.  Anyways, we have gone some distance towards
justifying the claims that the number of flux vacua is large, and that
it is useful to study their distribution statistically.

\bigskip

We thank B. Acharya, F. Denef, B. Doyon, P. Fonseca, G. Moore,
S. Murthy, P. Sarnak, B. Shiffman, S. Trivedi and S. Zelditch for
useful conversations and correspondence.  We particularly thank
F. Denef and G. Moore for numerous critical remarks, which are
addressed in the revised version.

This research was supported in part by DOE grant DE-FG02-96ER40959.

\listrefs
\end